\documentclass[manuscript]{aastex}
\usepackage{amssymb}
\usepackage{multirow}
\usepackage{rotating}
\usepackage{ctable}
\usepackage{morefloats}
\usepackage{sidecap}
\usepackage[colorlinks,breaklinks,citecolor=blue]{hyperref}
\usepackage{graphicx,epsfig,fancyhdr,psfig,rotating,amsmath,epsf,txfonts,natbib,epstopdf,multirow,appendix}

\def \halpha {H$\alpha$}
\def \kms {{ \rm km\;s$^{-1}$}}

\def \arcsec {$^{''}$}
\def \siiv {Si\,{\sc iv}}
\def \cii {C\,{\sc ii}}
\def \mgiik {Mg\,{\sc ii}\,k}
\begin{document}
\title{Explosive events on sub-arcsecond scale in IRIS observations: a case study}
\author{Zhenghua Huang\altaffilmark{1\bigstar},
Maria S. Madjarska\altaffilmark{2},
Lidong Xia\altaffilmark{1},
J.~G. Doyle\altaffilmark{2},
Klaus Galsgaard\altaffilmark{3},
Hui Fu\altaffilmark{1}
}

\altaffiltext{1}{Shandong Provincial Key Laboratory of Optical Astronomy and Solar-Terrestrial Environment, Institute of Space Sciences, Shandong University, Weihai, 264209 Shandong, China}
\altaffiltext{2}{Armagh Observatory, College Hill, Armagh BT61 9DG, N. Ireland}
\altaffiltext{3}{Niels Bohr Institute, 2100 Copenhagen, Denmark}
\altaffiltext{$\bigstar$}{Correspondence addressed to huangzhenghua@gmail.com}
 
\date{Received date, accepted date}

\begin{abstract}

We present study of a typical explosive event (EE) at sub-arcsecond scale witnessed by strong non-Gaussian profiles with blue- and red-shifted emission of up to 150~\kms\  seen in the transition-region \siiv\ 1402.8\,\AA, and the chromospheric  \mgiik\ 2796.4\,\AA\ and \cii\ 1334.5\,\AA observed by the Interface Region Imaging Spectrograph at unprecedented spatial and spectral resolution.  For the first time a EE is found to be associated with very small-scale ($\sim$120~km wide) plasma ejection followed by retraction in the chromosphere. These small-scale jets originate from a compact bright-point-like structure of $\sim$1.5\arcsec\  size as seen in the IRIS 1330~\AA\ images. SDO/AIA and SDO/HMI co-observations show that the EE lies in the footpoint of a complex loop-like brightening system. The EE is detected in the higher temperature channels of AIA 171~\AA, 193~\AA\ and 131~\AA\ suggesting that it reaches a higher temperature of log\,T\,$=5.36\pm0.06$\,(K). Brightenings observed in the AIA channels with durations 90--120 seconds are probably caused by the plasma ejections seen in the chromosphere. The wings of the \cii\ line behave in a similar manner as the \siiv's indicating close formation temperatures, while the \mgiik\ wings show additional Doppler-shifted emission.  Magnetic convergence or emergence followed by cancellation at a rate of $5\times10^{14}$\,Mx\,s$^{-1}$ is associated with the EE region. The combined changes of the locations and the flux of different magnetic patches suggest that magnetic reconnection must have taken place. Our results challenge several theories  put forward in the past to explain non-Gaussian line profiles, i.e. EEs. Our case study on its own, however, cannot reject these theories, thus further in-depth studies on the phenomena producing EEs are required.

 \end{abstract}


\maketitle

\section{Introduction}
\label{sect_intro}
The solar transition region is the interface between the chromosphere and the corona within which the temperature rapidly rises from 25\,000 K to 1 MK. Plasma in the solar transition region appears very dynamic evidenced by the so-called ``explosive events'' (EEs). The term `explosive event'  describes non-Gaussian (mostly transition region) line profiles  showing Doppler velocities of 50--150\,\kms\,\citep{1983ApJ...272..329B}. On average the EE size as determined along a spectrometer slit is  about 2\arcsec--5\arcsec\  with a lifetime of up to 600\,s\,\citep{1989SoPh..123...41D}. From 82 EEs observed by HRTS, \citet{1989SoPh..123...41D}  found only one case in which there was evidence for apparent velocities. This result suggests that the velocities of events associated with EEs are non-isotropic, and  (some or all) EEs are possibly the spectral signature of jets. EEs are often observed in bursts lasting up to 30 minutes \citep{1997SoPh..175..341I, Doyle2006}.

\par
EEs are usually found along the magnetic network at the boundaries of the super-granulation cells\,\citep{1989SoPh..123...41D,1991ApJ...370..775P, 2003A&A...403..731M}. They are associated with regions of weak and mixed polarity fluxes \,\citep{1988ApJ...335..986B,1991JGR....96.9399D,1998ApJ...497L.109C,2004A&A...427.1065T,2008ApJ...687.1398M}. \citet{1998ApJ...497L.109C} studied the magnetic  field of 163 EEs identified in Solar Ultraviolet Measurement of Emitted Radiation (SUMER) observations and Big Bear Solar Observatory (BBSO) magnetograms, and found that 103 of these were associated with magnetic flux cancellations. However, the connection between EEs and magnetic cancellation is still under debate. \citet{2008ApJ...687.1398M} found that only 7 out of 37 EEs were associated with magnetic cancellation sites while the magnetic flux for   62\% of EEs did not change during their lifetime, though it is possible that this is due to instrumental limitations. Magnetic reconnection is proposed as the possible mechanism that produces opposite directed  jets generating  the EE's blue- and red-shifted emission \citep[see e.g.][]{1991JGR....96.9399D, 1997Natur.386..811I, 1998ApJ...497L.109C,2000ApJ...541L..29R,2000ApJ...545.1124L}.

\par
The true nature of the events associated with EEs remains unknown as these `events' actually  carry  only the spectral signature about the observed phenomena. EEs were suggested to be the signature of siphon flows in  small-scale loops \citep{2004A&A...427.1065T}.  They are also believed to be produced by spicules and macrospicules  \citep{2000A&A...360..351W}, and were found to be associated with chromospheric upflow events \citep{1998ApJ...504L.123C}.  EEs were found in transient brightenings and X-ray jets \citep{2012A&A...545A..67M}. \citet{2009ApJ...701..253M} showed that EEs can result  from up- and down-flows in a surge. \citet{2011A&A...532L...9C} put forward the idea that EEs are produced by swirling jets where a helical motion would be mostly responsible for the blue- and red-shifted  emission (the Si~{\sc iii} 1206.51~\AA\ line was used in this study).

 \par
EEs are typically observed in transition region emission lines with formation temperatures ranging from $2\times10^4$\,K to $5\times10^5$\,K  \citep{1983ApJ...272..329B}. \citet{1992sws..coll...11D} reports from HRTS spectra that less than 1\% of the EEs observed in transition region lines are also seen in {\sc C\,i}\,1561\,\AA\,($1\times10^4$\,K) while they are weakly seen in {\sc C\,ii}\,1335\,\AA\,($1.6\times10^4$\,K). SUMER observations showed  that EEs  also appear in lower temperature lines such as {\sc O\,i} ($1\times10^4$\,K), Lyman\,6 to Lyman\,11\,\citep[$1.2\times10^4$\,K,][]{2002A&A...382..319M}, and Lyman\,$\beta$\,\,\citep[$1.2\times10^4$\,K,][]{2010A&A...520A..37Z}. These lines have a non-Gaussian shape with a reversed line core surrounded by two emission peaks. In Lyman 6--11, \citet{2002A&A...382..319M} found that some of these lines show a stronger self-absorption in EE which is due to an increase of the emission in the wings while the core intensity increase remains weak. In Lyman\,$\beta$, \citet{2010A&A...520A..37Z} found that the self-reversion becomes more significant during EEs with a stronger peak in the blue wing.
 
 \par
 \citet{1998ESASP.421..103W} identified two EEs that occurred  in an active region and were observed at coronal temperatures (Mg\,{\sc ix}\,749\,\AA, $10^6$\,K). \citet{2002A&A...392..309T} found that EEs in the quiet Sun do not have a coronal response suggesting that they are not relevant to coronal heating (only 2 events were analysed in this study). From the analysis of the energetics of explosive events observed with SUMER, \citet{2002ApJ...565.1298W} concluded that the energy released in explosive events should be enough to heat the solar atmosphere.

\par
The overview given above shows that our knowledge of the phenomena generating EEs is very uncertain with many open questions which need to be addressed, e.g., what physical phenomena generate EEs' up- and down-flows, rotation or all simultaneously? Do phenomena associated with EEs contribute directly or indirectly to the mass and energy transfer in the solar atmosphere? What is(are) the physical  mechanism(s) driving the phenomena that produce explosive events? Where in the solar atmosphere do phenomena  producing explosive-event line profiles typically originate? The combination of the very recent state-of-the-art mission, the Interface Region Imaging Spectrograph explorer (IRIS), with the excellency of the Atmospheric Imaging Assembly (AIA) and the Helioseismic and Magnetic Imager (HMI) data, provide an unprecedented opportunity that might help us answering the many open questions related to EEs. Our study does not give an answer of all the open questions. However, it provides the first insight  into the unique capabilities of IRIS (spectral and imaging chromospheric and transition-region data  with remarkable spatial and spectral resolution) for  unlocking the mystery of what phenomena drive the feature called ``explosive event".

\par
In the present work, we carry out a case study of an explosive event combining simultaneous IRIS spectroscopic and imaging data together with magnetic field and coronal imaging observations from HMI and AIA onboard the Solar Dynamics Observatory (SDO). We report for the first time and in unprecedented detail on the spectral and imaging characteristics of an explosive event on a sub-arcsecond scale which should give  a better understanding on the driving physical mechanism of these phenomena.  We also describe the behavior of the observed optically thick lines which is crucial for a radiative transfer modelling of the emission from dynamically evolving atmospheric phenomena. This study will promote an extensive statistical study based on spectral (IRIS) and imaging (IRIS \& AIA) data that will be crucial in making a step forward into the understanding of the solar phenomena that trigger strong non-Gaussian profiles in the solar chromosphere and transition region, i.e.``explosive events'', and their true role in coronal heating. The article is organised as follows:  Section\,\ref{sect_obs} describes the observations and the data reduction, the results and discussion are presented in Section\,\ref{sect_res}, and the conclusions are given in Section\,\ref{sect_concl}.

\section{Observations and data reduction} 
\label{sect_obs}

The observations were taken by the IRIS\,\citep{2014SoPh..tmp...25D}, the AIA\,\citep{2012SoPh..275...17L} and the HMI\,\citep{2012SoPh..275..229S} onboard SDO\,\citep{2012SoPh..275....3P} on 2013 October 4 from 18:42:09\,UT to 19:00:54\,UT.

\subsection{IRIS}

\begin{figure*}
\includegraphics[trim=1cm 0.5cm 0.5cm 0cm,clip,width=17cm]{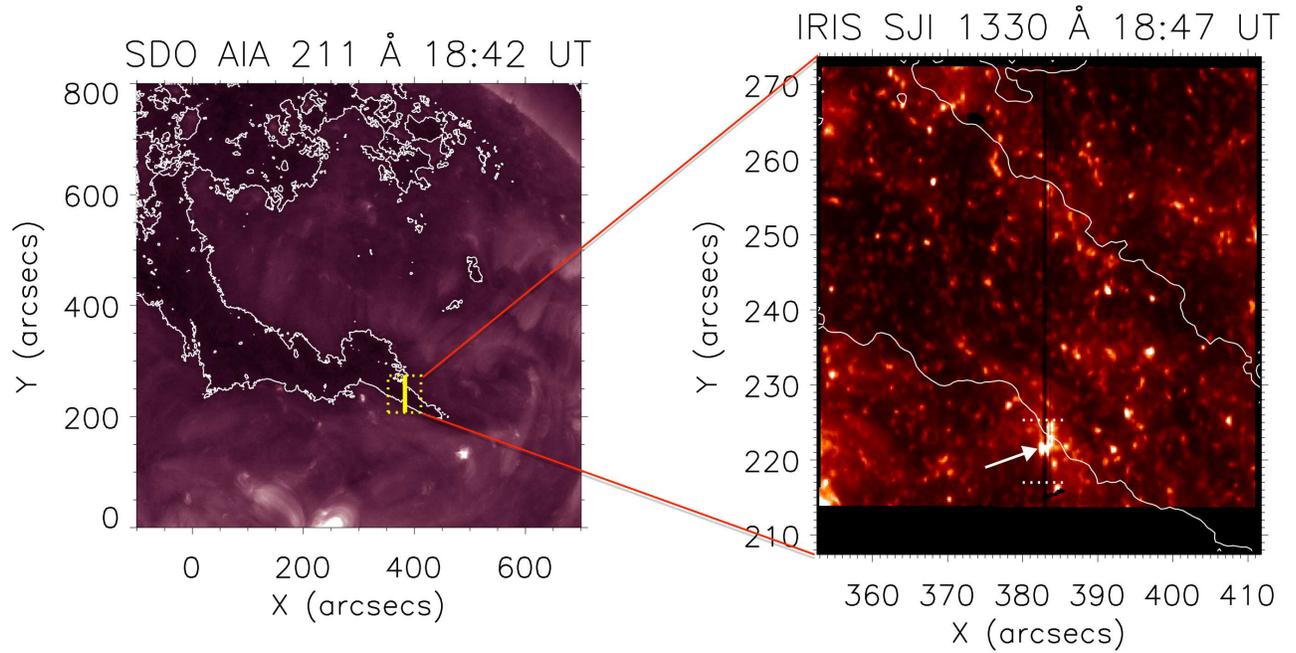}
\caption{Left: Cut-off AIA 211\,\AA\ image with  over-plotted the contour of the equatorial extension of a polar coronal hole. The outlined dotted-box region is  the IRIS slit-jaw image field-of-view. The vertical line represents the IRIS slit position. Right: IRIS slit-jaw image on which the dark vertical line in the middle is the location of the slit. The  contour plot is the boundary of the coronal hole defined in the AIA 211~\AA\ image. 
\label{fig_fovpos}}
\end{figure*}

IRIS is a NASA Small Explorer mission, which was launched into a sun-synchronous orbit on 27 June 2013.   It takes simultaneous spectra and images of the interface region (i.e. chromosphere and transition region) of the Sun with 0.33\arcsec -- 0.4\arcsec\ spatial resolution and as low as  2\,s cadence. IRIS observes the far UV wavelength band in  the  spectral range from 1331.6 to 1358.4~\AA\ that includes two bright {\sc C ii} lines with a spectral sampling of 12.98~m\AA/pixel,   and in  the range 1380.6--1406.8~\AA\ (containing several Si\,{\sc iv} and O\,{\sc iv} lines) at 12.72~m\AA/pixel. IRIS also operates in the near UV spectral range from 2782.6 to 2833.9~\AA\  at 25.46~m\AA/pixel recording  two bright Mg\,{\sc ii} lines. For comparison, the SUMER spectral sampling is around 44.0--45.0~m\AA/pixel in the 660--1600~\AA, i.e. IRIS has more than three times higher spectral resolution. The instrument is designed in such a way that a slit with a size 0.33\arcsec\ $\times$ 175\arcsec\ guides the sunlight into a spectrograph while a reflective coating directs the sunlight outside of the slit into an imaging system making slit-jaw images (SJIs). Therefore, the slit position appears as an emission-blocked vertical line on the slit-jaw images (see the example in Fig.\,\ref{fig_fovpos}). The slit can either stay in a fixed position to record sit-and-stare spectra or scan a large area with a selected step size.
 The slit-jaw images can be taken in four different channels, one centred at 1330\,\AA\ with a 40\,\AA\ bandpass recording chromospheric and strong continuum  emission, one at 1400\,\AA\ again with a 40\,\AA\ bandpass imaging the lower transition region, one at 2796\,\AA\ with a 4\,\AA\ bandpass providing images of the upper chromosphere, and one at 2832\,\AA\ with a 4\,\AA\ bandpass for high-contrast photospheric imaging.

\par
The IRIS dataset used in this study includes spectral observations taken in a sit-and-stare mode with 8\,s exposure time and 9\,s cadence, and slit-jaw images taken in the 1330\,\AA\ channel also at  9\,s cadence. Four spectral windows were transferred to the ground containing two C\,{\sc ii} lines (1334.5\,\AA\ and 1335.7\,\AA, $\log$T=4.3\,K), the Si\,{\sc iv} 1402.8\,\AA\ line ($\log$T=4.8\,K) and the \mgiik\ 2796.4\,\AA\ line ($\log$T=4\,K). IRIS was targeting an equatorial extension of a polar coronal hole across its boundaries (Fig.\,\ref{fig_fovpos}). Fig.\,\ref{fig_fovpos} shows an overview of the coronal hole in the 211~\AA\ channel of AIA and one of the SJ images.  The coronal hole boundaries are defined as the region where the AIA 211\,\AA\ emission drops to half of that in the surrounding quiet Sun region. The spectral observations were taken in a sit-and-stare mode, but a compensation for the solar differential rotation was not applied to the data.

\par
The present study uses IRIS level 2 data provided by the IRIS team. These are science products after dark current removal, flat-field and geometric correction have been applied. We found that spiked pixels are still present in the spectral data. To flag these pixels, we first determined the maxima of  the line profiles and the standard deviation ($\sigma$) of the maxima were calculated. The pixels with line profiles that have a maximum exceeding  $3\sigma$ were flagged and excluded from any further analysis. A further wavelength correction for the orbital variation of the line position was required due to the temperature change of the detector and the spacecraft--Sun distance variation over the course of an orbit. To correct this, we used the standard program provided by the IRIS team \citep{2014arXiv1404.6291T}.

\subsection{AIA and HMI}
The AIA observations used in this study were taken in the UV channels including the 1700\,\AA\ and 1600\,\AA\  at about 24\,s cadence, and in  the EUV 304\,\AA, 171\,\AA, 193\,\AA, 211\,\AA, 335\,\AA, 131\,\AA\ and 94\,\AA\ passbands at about 12~s cadence. The spatial resolution of the AIA images is 1.2\arcsec (0.6\arcsec per pixel). The HMI longitudinal magnetograms  analysed  here have a 45\,s cadence and were  taken from 17:40\,UT to 19:17\,UT (IRIS observed from 18:42~UT to 19:00~UT) in order to investigate the magnetic field evolution that precedes and follows the IRIS observations. The HMI pixel size is 0.505\arcsec\
and a 1$\sigma$ noise level is 10~G \citep{2012SoPh..279..295L}.

\subsection{Co-alignment}

The emission in the three IRIS spectral lines (\mgiik, C\,{\sc ii}, and Si\,{\sc iv}) is shifted along the IRIS slit.  To correct for this offset we used  fiducial emission marks made by various phenomena along the slit. We found an offset between  C\,{\sc ii} and  Si\,{\sc iv} of 7 pixels (i.e.  Si\,{\sc iv} is 7 pixels lower than \cii) and \mgiik\ is 4  pixels higher than  C~{\sc ii}.   The IRIS 1330\,\AA\ slit-jaw and  the AIA 1600\,\AA\ images both have a strong continuum contribution which permits a straightforward alignment between both datasets. The AIA 1600\,\AA\ images were then aligned with the HMI data. 

\section{Data analysis, results and discussion}
\label{sect_res}

The IRIS observations present a unique opportunity to study  the solar chromosphere and transition region at unprecedented spatial (sub-arcsecond), spectral  and time resolution. Here, we explore this advantage to analyse at sub-arcsecond scale a small (close to 1\arcsec size) transient phenomenon  combining  spectroscopic,  imaging and magnetic-field co-observations. The online animation (see Fig.~\ref{fig_movie1}) provides spatially and temporally combined observations from all the instruments used here, i.e. the SJ 1330~\AA\ images from IRIS, the HMI magnetograms, the AIA images in the 1600~\AA, 304~\AA, 171~\AA, 193~\AA, 211~\AA\ and  131~\AA\ channels together with the emission along the slit in the three  IRIS spectral lines, \mgiik, \cii\ and \siiv. The spectroscopic investigation and results are presented in Sect.~\ref{sect_ee_sp}, the imaging (IRIS and AIA) analysis is given in Sect.~\ref{sect_photo}, the magnetic field study is described in Sect.~\ref{sect_mag}. We also present our interpretation of the observed spectral and imaging features, the possible physical mechanism and  the challenges brought up by the instrumental limitations. 

\subsection{IRIS spectroscopy of the explosive event}
\label{sect_ee_sp}
\begin{figure*}
\includegraphics[trim=0.5cm 0cm 0cm 0cm,clip,width=17cm]{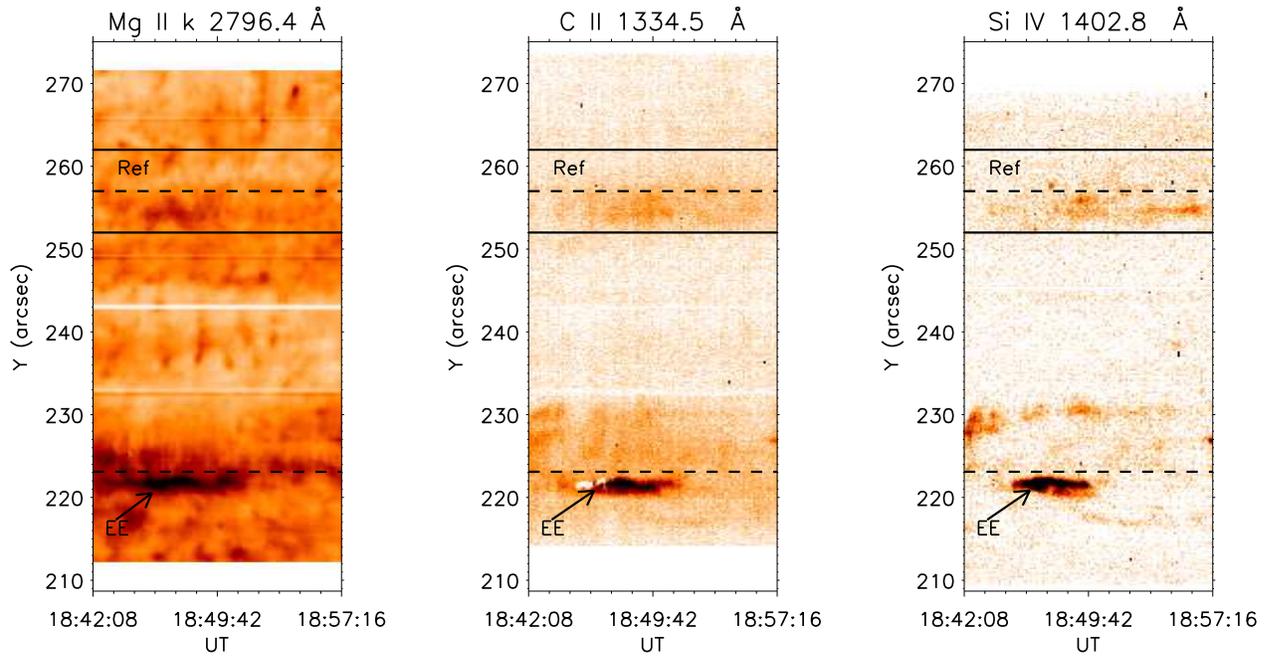}
\caption{Radiance images in \mgiik\,2796.4\,\AA, C\,{\sc ii}\,1334.5\,\AA\ and Si\,{\sc iv}\,1402.8\,\AA\ (in reversed colour). The coronal hole boundary determined from  the AIA 211\,\AA\ image is marked with a dashed line. The region between the two solid lines  marked with `Ref' is where the reference line profiles are obtained. An  arrow with `EE' points at the studied explosive event .\label{fig_iris_sp_imgs}}
\end{figure*}

The IRIS spectral and imaging data were taken at the boundaries of an equatorial extension of a polar coronal hole (Fig.~\ref{fig_fovpos}).  The spectral observations obtained in the \siiv\ line were first analysed by applying a single Gaussian fit. Then the dataset was inspected for strong blue or/and red-shifted emission above 50~\kms. We identified one such event with  strong non-Gaussian profiles (solar\_y=200\arcsec). Judging by its spectroscopic appearance, the event carries all the observational characteristics of the phenomenon called `explosive event' (EE, see Section\,\ref{sect_intro}  for more details). The  EE measures  $\approx$1.5\arcsec along the IRIS slit.  In the radiance images, produced from the sit-and-stare slit observations, it  appears  as a bright compact structure in the \mgiik, \cii\ and \siiv\  lines (Fig.\,\ref{fig_iris_sp_imgs}). Note that because compensation for the solar differential rotation  was not applied during the IRIS observations, at this heliographic coordinates the slit scans an area of 1\arcsec\ in 460~s. For an event with a size of 1.5\arcsec, it will take approximately 11~min to be scanned from west to east. In the SJ images, the EE appears as a compact bright-point-like structure which is visible during the whole observing period, i.e. 18~min 46~s.

\begin{figure*}
\includegraphics[width=\textwidth,clip,trim=0cm 1cm 0.5cm 0cm]{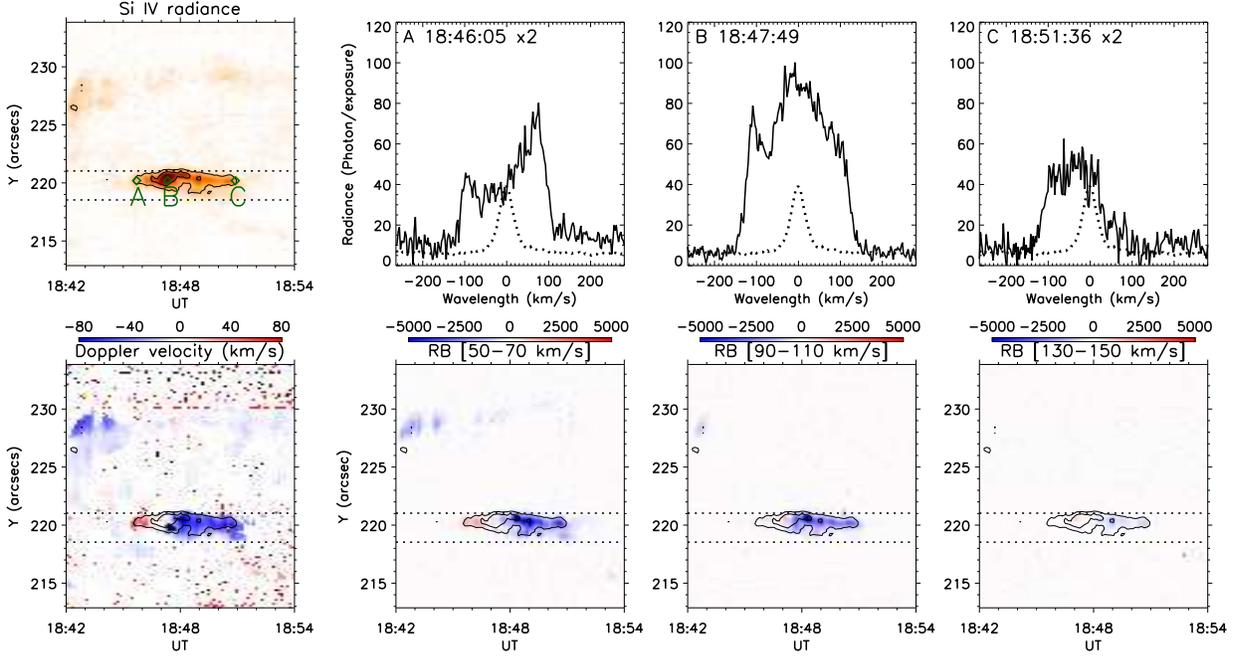}
\caption{The explosive event in  \siiv. {\bf From left to right, top row}: radiance image (in reversed colour) produced from the total intensity of the \siiv\ line, \siiv\ profiles taken from the pixels denoted with diamond symbols as `A', `B', and `C' in the radiance image.The dotted line, multiplied by 5, is the reference spectrum obtained from the region shown in Fig.~\ref{fig_iris_sp_imgs}. The \siiv\ profiles from `A' and `C' are multiplied by 2. {\bf Bottom row}: Doppler velocity image derived from a single Gaussian fit, RB asymmetry of the region of the explosive event at 50--70\,\kms, 90--110\,\kms, and 130--150\,\kms\ with over-plotted  the contour of the radiance image. The dotted lines outline the region from which the temporal variations of RB (Fig.\,\ref{fig_rb_si4}) are obtained.  \label{fig_ee_si4}}
\end{figure*}

In Fig.\,\ref{fig_ee_si4} (top and bottom left) we show the radiance  and the Doppler-shift images obtained from  the \siiv\ line. The Doppler-velocity image was produced by applying a  single Gaussian fit. Away from the EE, the line is dominated by noise due to the low count rate of \siiv\ but the emission in the EE has a good signal-to-noise ratio allowing reliable calculation from a  single Gaussian fit. As we mentioned above  the spectral information obtained during the event is  both spatial and temporal, although the observations are in a sit-and-stare mode. While the EE moves under the IRIS slit, first a red-shift dominated emission is registered for the time between 18:45:27\,UT and 18:46:42~UT, i.e. during 75~s.  The Sun  would rotate  under the IRIS slit  for this period of time by approximately 0.16\arcsec, i.e. $\sim$118~km.  An example of the red-shifted \siiv\ line profile is given in Fig.~\ref{fig_ee_si4} (top row, 2nd panel--``A''). Gradually blue-shifted emission is seen to increase  from 18:46:57~UT to 18:47:49~UT, (i.e. in 52~s) until  the emission in both wings is almost equal   (Fig.~\ref{fig_ee_si4}, top row, 3nd panel--``B'').  From 18:47:58~UT until 18:51:55~UT (237~s), the emission is blue-shift dominated (Fig.\,\ref{fig_ee_si4}, top row, 4th panel--``C''). The phenomenon was scanned in 388~s which means that the size of the EE in the \siiv\ line (i.e., transition region) is $\approx$0.84\arcsec. Considering the projection angle, the EE should have a size of 0.94\arcsec\ on the solar surface.

\par
The Doppler-shift pattern seen in the single Gaussian images was further  investigated using the red-blue (RB) asymmetry method. RB asymmetry is defined as the difference of the emission in the red and blue wing of a spectral line at the same Doppler velocity (or Doppler-velocity range) and is given by:
$$RB_{\Delta\lambda_1}=\int_{\lambda_0+\Delta\lambda_1-\delta\lambda_{\omega}}^{\lambda_0+\Delta\lambda_1+\delta\lambda_{\omega}}I_\lambda d\lambda - \int_{\lambda_0-\Delta\lambda_1-\delta\lambda_{\omega}}^{\lambda_0-\Delta\lambda_1+\delta\lambda_{\omega}}I_\lambda d\lambda,$$
where $\lambda_0$ is the wavelength of the line centre, $I_\lambda$ is the intensity of the spectral line,  $\Delta\lambda_1$ is the offset from the line centre and $\delta\lambda_\omega$ is  the wavelength range over which the RB asymmetry is determined. This method has been widely used in optically thick spectral-line analysis \citep[e.g., in \halpha,][and the references therein]{2009ApJ...701..253M,huangzh2013}, and it was also introduced to emission lines \citep{2009ApJ...701L...1D}. A variant of this method that differs in the determination of the line centre is  discussed in \citet{2011ApJ...738...18T}. In the present study, the line centre is obtained from an average profile in the region indicated in Fig.\,\ref{fig_iris_sp_imgs}.  In Fig.~\ref{fig_ee_si4}, bottom row, we show images of  the RB asymmetry in the \siiv\ line in three Doppler-shift ranges: 50--70~\kms, 90--110~\kms\ and 130--150~\kms. The temporal variations of the RB asymmetry obtained by averaging over the EE along the slit in the \siiv\ line are given in Fig.~\ref{fig_rb_si4}, top panel.  We clearly see from both figures that the RB asymmetry of the EE is positive only in the lowest velocity range (50--70~\kms). The blue-shift  dominant emission is found in the central part of the EE. Again we have to keep  in mind that the information is space-time combined  and the excess up-flow may later be followed by a down-flow or vice-versa. As seen by \citet{2009ApJ...701..253M} in the case of a surge,  simultaneous plasma up- and down-flow along adjacent  field-lines would produce simultaneous blue and/or red-shifted emission with the appearance along the spectrometer slit depending on the line-of-sight.

\par
The observations show that the red-shifted emission is weaker than the blue-shifted indicating that less plasma at the formation temperatures of the three lines falls back towards the chromosphere.  One explanation could be that the plasma is ejected from the chromosphere and part of it is heated to transition-region or/and coronal  temperatures. After the plasma  deposits  the heat in the upper atmosphere, it falls back to the chromosphere at different speeds and over a longer period of time. Another possible interpretation  is that the plasma is ejected along looped magnetic-field  lines thus leaving the field-of-view of the spectrometer slit.
\par
\begin{figure*}
\includegraphics[width=\textwidth,clip,trim=0cm 1cm 0.5cm 0cm]{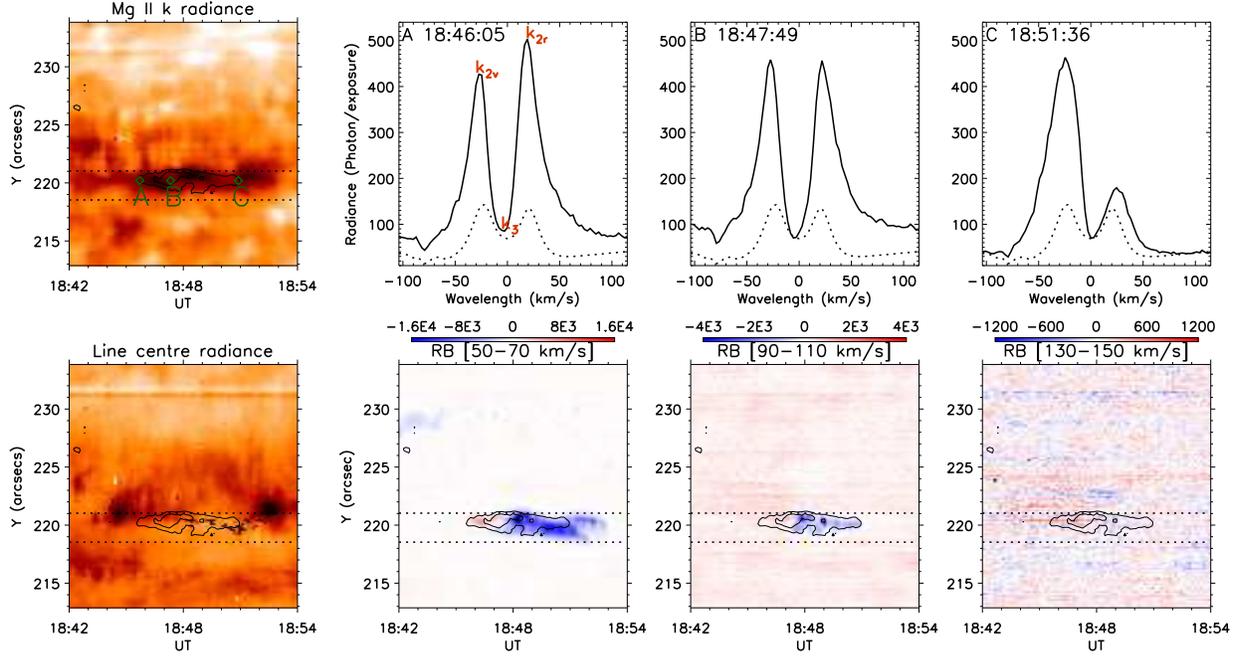}
\caption{The explosive event in IRIS \mgiik. {\bf From left to right, top row}: radiance image (in reversed colour) from the total intensity in the \mgiik\ line,  \mgiik\ line profiles  taken from the pixels denoted by diamond symbols  as `A', `B', and `C', together with over-plotted  the reference profile (dotted line). The blue emission peak is marked with  `k$_{2v}$', the red --  `k$_{2r}$' and the absorbtion core as `k$_{3}$'. {\bf Bottom row}: radiance image (in reversed colour) in the line centre of \mgiik\ (i.e. in k$_{3}$), RB asymmetry of the region of the explosive event at 50--70\,\kms, 90--110\,\kms, and 130--150\,\kms\ with over-plotted the contour of the radiance image. The dotted lines outline the region from which the temporal variations of RB (Fig.\,\ref{fig_rb_si4}) are obtained.
\label{fig_ee_mg2k}}
\end{figure*}
\begin{figure*}
\includegraphics[width=16cm,clip,trim=0.5cm 1cm 0.5cm 0cm]{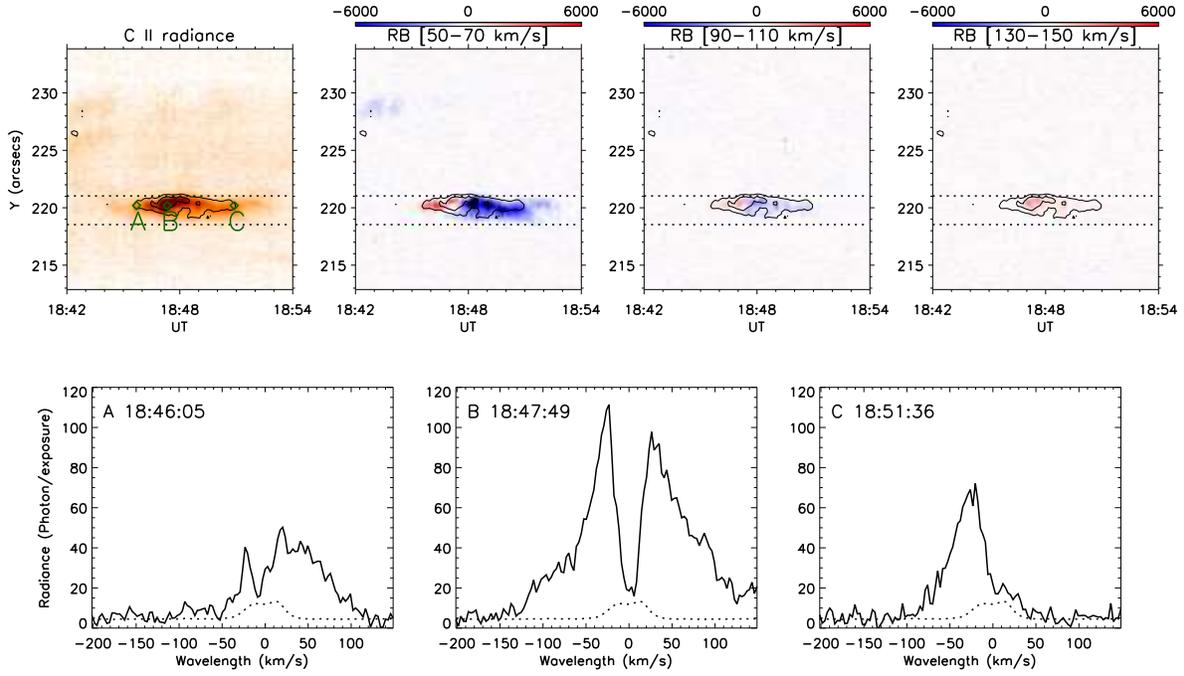}
\caption{The explosive event in  \cii\,1334.5\,\AA. {\bf Top row from left to right}: radiance image (reversed colour), RB asymmetry of the region of the explosive event at 50--70\,\kms, 90--110\,\kms\ and and 130--150\,\kms. The dotted lines outline the region from which the temporal variations of RB (Fig.\,\ref{fig_rb_si4}) are obatined. The contours from the \siiv\ radiance image is over-plotted on all images. Bottom row from left to right:  \cii\,1334.5\,\AA\ profile (solid lines) taken from pixels `A', `B', and `C' with over-plotted the reference profile (dotted line). \label{fig_ee_c2}}
\end{figure*}
\begin{figure*}
\includegraphics[width=8cm,clip,trim=0.5cm 3cm 0cm 0.8cm]{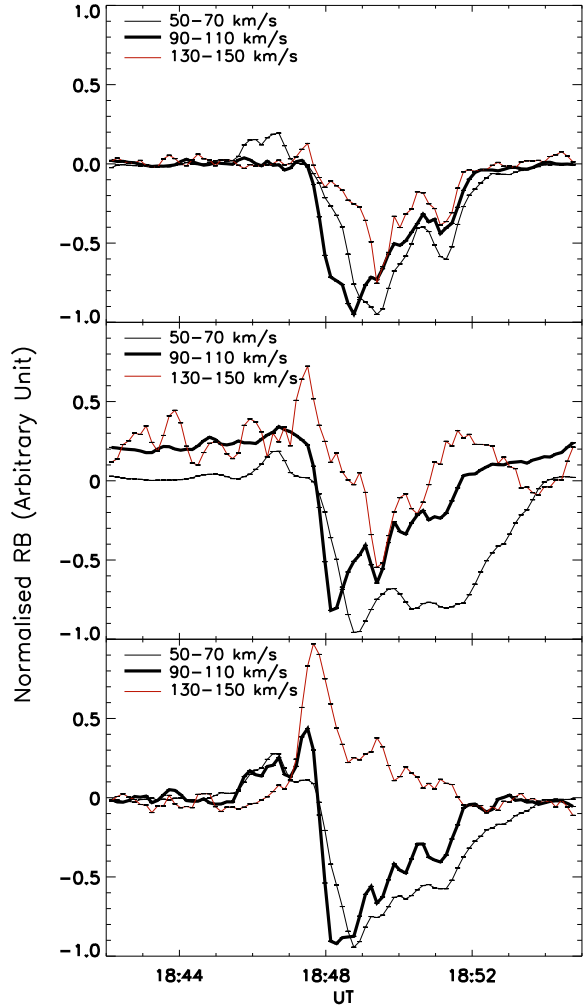}
\caption{Temporal variations of the normalised RB asymmetry in the three different Doppler-shift ranges from the region outlined with dotted lines in Figs.~\ref{fig_ee_si4}--\ref{fig_ee_c2}. {\bf From top to bottom}: \siiv, \mgiik\ and \cii. \label{fig_rb_si4}}
\end{figure*}

The event also appears as a compact bright structure in the \mgiik\ and \cii\ radiance images (Figs.\,\ref{fig_ee_mg2k} and \ref{fig_ee_c2}) similar to \siiv.   Both lines appear with  a central absorption core surrounded by two emission peaks. The \mgiik\ blue peak is sometimes referred  to as k$_{2v}$ and the red as k$_{2r}$, while the line core -- k$_{3}$.  Here, we will simply refer  to k$_{2v}$  and k$_{2r}$ as blue and red wings. 
The line profiles of the two absorption lines at three sampling pixels (``A'', ``B'' and ``C'' in bottom rows of Figs.\,\ref{fig_ee_mg2k} and \ref{fig_ee_c2}) are red-wing-dominated, with equal wings, and blue-wing-dominated respectively, i.e. the same as  for the \siiv\ profiles. The RB asymmetry  in the two chromospheric lines is derived  as for the \siiv\ line and is  shown in Figs.\,\ref{fig_ee_mg2k} and \ref{fig_ee_c2}.  The RB temporal variations are given in Fig.\,\ref{fig_rb_si4}. Note that Doppler-shifts  in the wings of the \mgiik\ line are reported to be a signature of flow velocities as found by \citet{Leenaarts2013a,Leenaarts2013b} and \citet{Pereira2013}.  The \mgiik\ and \cii\ RB asymmetry images  show very similar to behaviour as the \siiv\ asymmetry with the only difference in the Doppler velocity range 130--150\,\kms. That may be  due to a blend in the red wing of these two lines (\mgiik\ is  blended by an unidentified line, and \cii\,1334.5\,\AA\ is blended by the blue wing of \cii\,1335.7\,\AA). Our results demonstrate that above 50~\kms\ during a very dynamic event the wings of the chromospheric lines studied here carry identical information as the optically thin \siiv\ line. 

 \par
\begin{figure*}
\includegraphics[width=15cm,clip,trim=1cm 3cm 0.5cm 0.8cm]{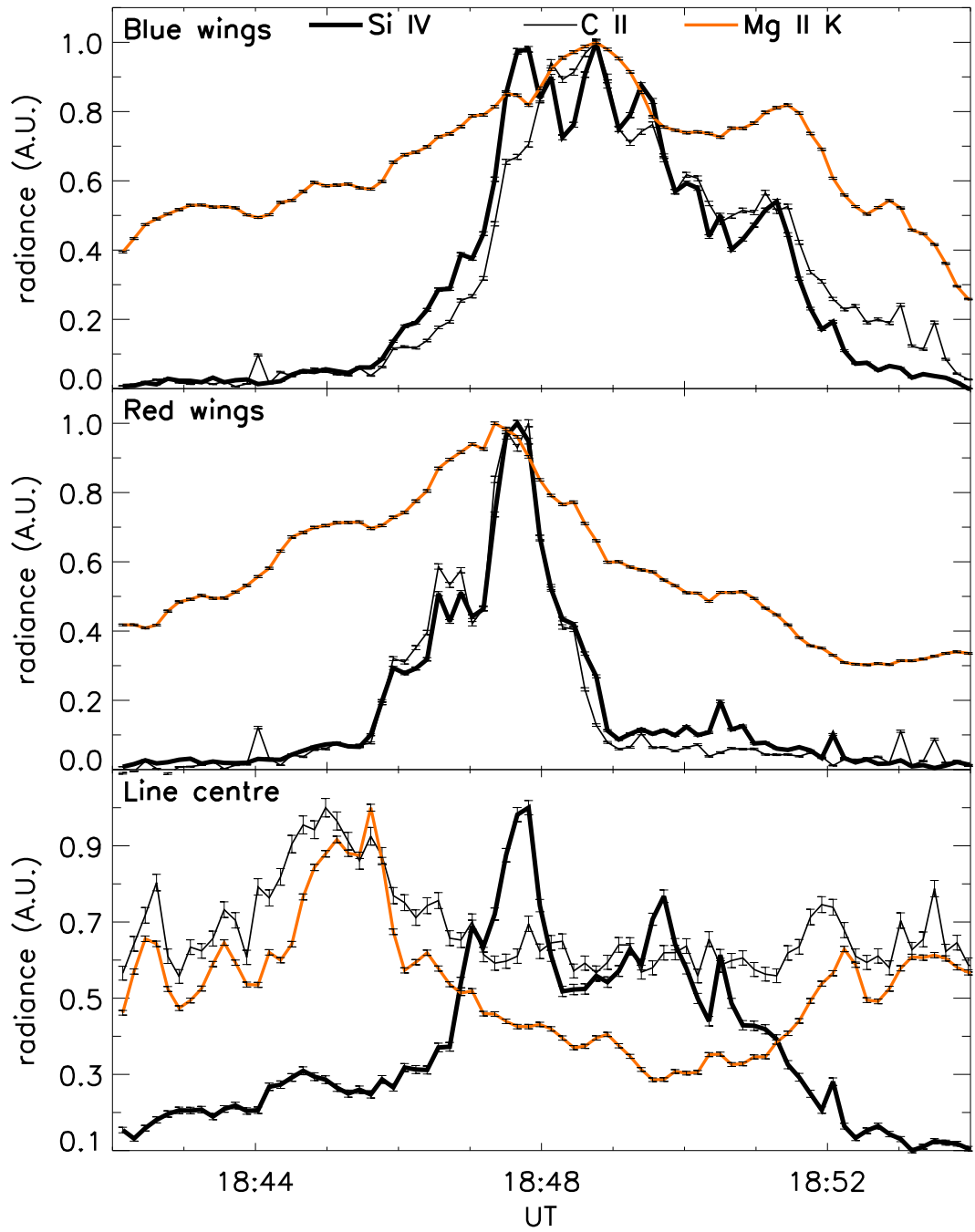}
\caption{Lightcurves  in the blue (top) and red (middle) wings of \siiv, \cii, and \mgiik\ in the range 20$\--$100 \kms, and the line centre from $-$5\,\kms\ to 5\,\kms\ (for all three lines, bottom). Please note that the slit was crossing the event from west to east (see Sect.~\ref{sect_ee_sp} for more details).\label{fig_ee_spws_lcs}}
\end{figure*}

To investigate in more detail  the radiance behavior in the  wings and centres of the three lines, we produced lightcurves in their  blue and red wings and  line centres (Fig.~\ref{fig_ee_spws_lcs}). The response in the wings of \cii\ is almost identical  to  the \siiv\ wing emission with only a small delay of 9~s for the blue wing of \cii\  to reach peak emission (top panel in Fig.~\ref{fig_ee_spws_lcs}).   This indicates that the wings of the two lines form at a similar temperature. The behavior of the  \mgiik\ wings is similar to that of the \cii. However, the emission in the wings of \mgiik\  in addition to the Doppler shift is possibly also affected by  emission  related to wing  formation in the lower chromosphere.  The above is not valid, however, for the line centres. The line centre of \cii\ is clearly reversed. We obtained  the lightcurves in the line centre  of the \mgiik\ and \cii\ lines   summing the emission from $-$5 to 5~\kms. The line-centre emission of these two lines  show the same behavior  which is very different from the emission in the centre of the \siiv\ line  (Fig.~\ref{fig_ee_spws_lcs}, bottom panel). The only difference between  \cii\ and \mgiik\ is the later response of \mgiik\ (18~s) when a sudden intensity increase at the edges of the brightening linked to the EE is observed. This clearly shows that the line centres of both chromospheric lines are emitted from plasma at similar temperatures. 

\par
A line centre reversal  in the \cii\ has been seen in HRTS (0.05~\AA\ spectral resolution) data but has never been reported in SUMER  (0.045~\AA) observations \citep[Figs.~9 and 10 in][]{2013ApJ...779..155A}. We made an automatic scan of  all the \cii\ data in  the SUMER archive visually searching for \cii\ profiles showing explosive  event line profiles. We found numerous examples of EEs where \cii\ shows a central reversal but only in stronger events. One of the reasons that a central reversal is now clearly seen in the IRIS data is apparently due to the IRIS higher spatial  resolution. In SUMER, a small central reversal will not be visible but instead a flat profile peak will be registered. 
 
 \par
\begin{figure*}
\includegraphics[width=8cm]{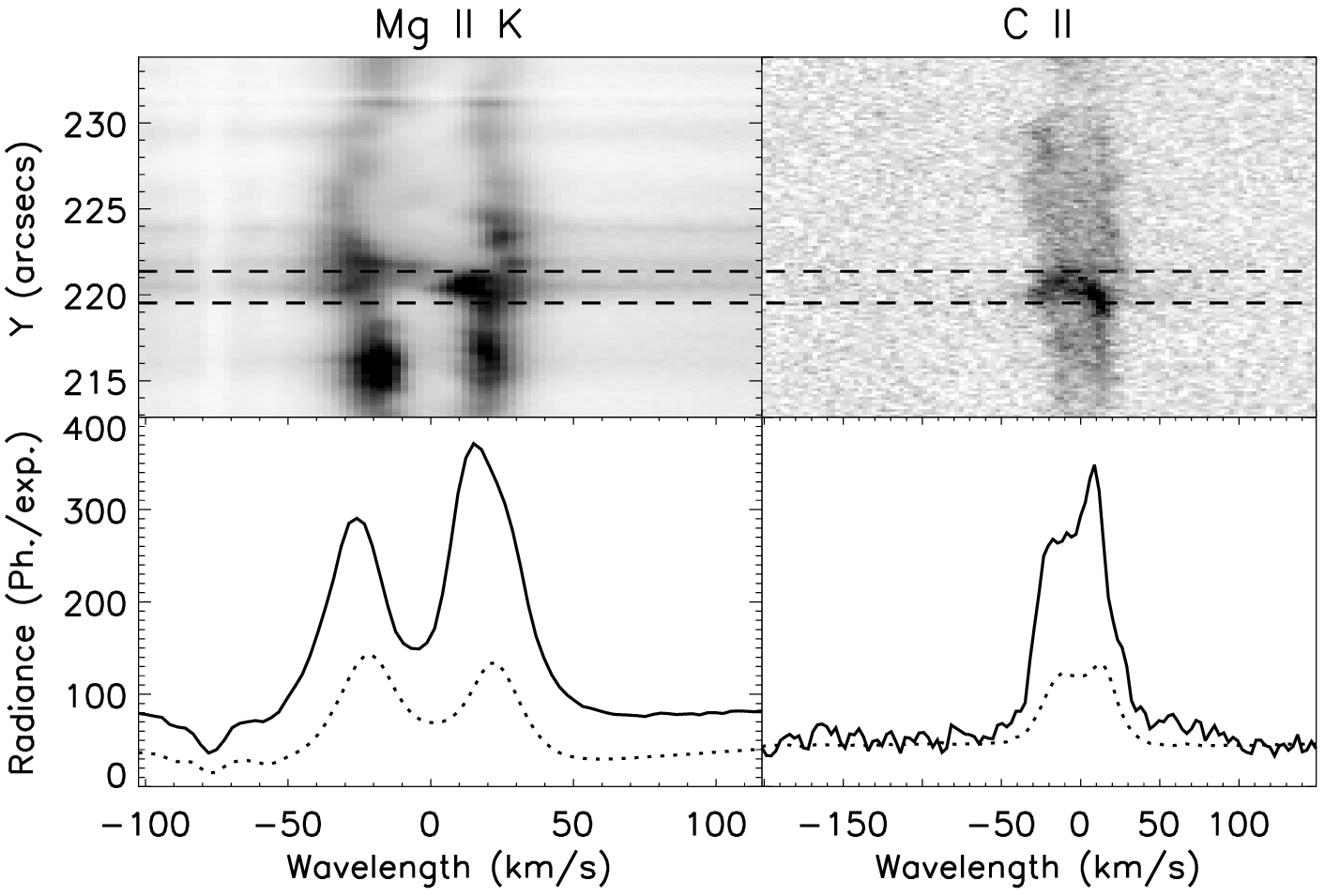}\\
\includegraphics[width=8cm]{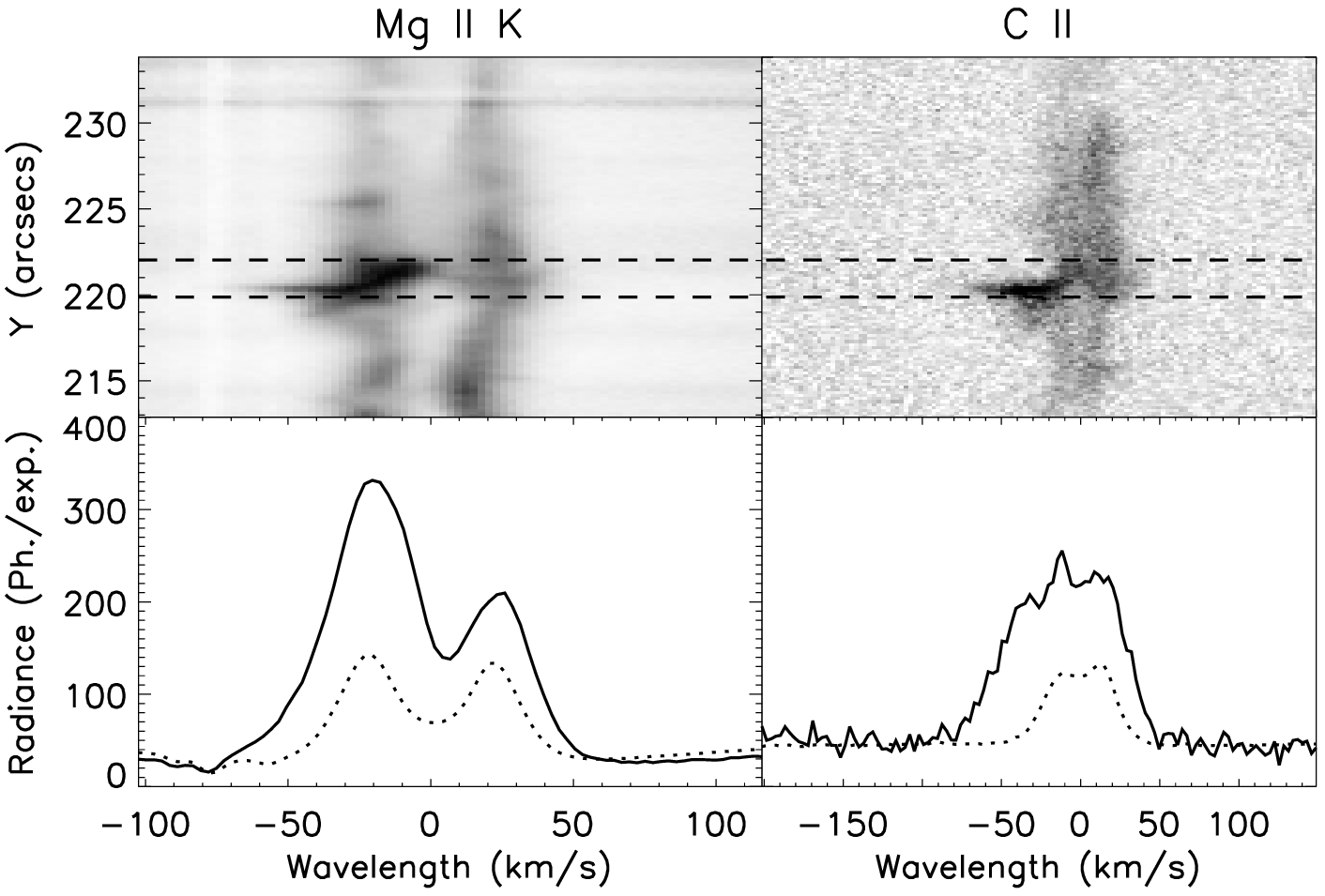}
\caption{\mgiik\ and \cii\ profiles taken from the regions where a line centre increase is observed (see the bottom-left panel of Fig.\,\ref{fig_ee_mg2k}). {\bf Top panel}: profiles taken at 18:44:58\,UT. {\bf Bottom panel}: profiles taken at 18:52:51\,UT. The dashed lines indicate the area from which the line profile plots  are produced. The \cii\ profiles are enlarged by a factor of 10 in order to fit them to the same radiance axis as \mgiik. The over-plotted dotted lines are the average line profiles of the `Ref' region  in Fig.\,\ref{fig_iris_sp_imgs}.}\label{fig_ee_cn_prof}
\end{figure*}

 An emission increase is clearly seen  in the line centre of  \mgiik\ and \cii\ at the west (the start of the scanning of the event as observed in a sit-and-stare mode) and east (end of the scanning) edge of the event. In Fig.~\ref{fig_ee_mg2k} (bottom row,  first panel), we show the  \mgiik\ line-centre radiance image where the two brightenings are seen. After careful investigation we found that this increase is not due to a line centre emission increase but rather to a shift of the wings of both the \cii\ and \mgiik\ lines towards the line centre. On the west  site of the event the intensity in the red wing increases and shifts towards the line centre (shorter wavelength) while on the other edge (east or end of the scanning) of the event the intensity in the blue wing  rises shifting again towards the line centre (longer wavelength) (Fig.~\ref{fig_ee_cn_prof}). This behavior appears temporally and spatially where  the \siiv\ line-wing-emission excess is observed, and it suggests that an intensity increase towards the line centre is related to the plasma dynamics. Radiative transfer calculations  are  only available under the assumption of ionisation equilibrium, therefore, these are not entirely valid for the atmospheric conditions in the presence of a highly energetic event \citep{Leenaarts2013a,Leenaarts2013b,Pereira2013}. \citet{2013A&A...557L...9D}  showed that for dynamic bursts with a decay time of a few seconds, the \siiv\ line can be enhanced by a factor of 2--4 in the first fraction of a second with the peak in the line contribution function occurring initially at a higher electron temperature due to transient ionisation compared to ionisation equilibrium conditions.

\subsection{IRIS and AIA imaging of the explosive event}
\label{sect_photo}
\begin{figure*}
\includegraphics[width=17cm,clip,trim=0cm 7cm 01cm 0.5cm]{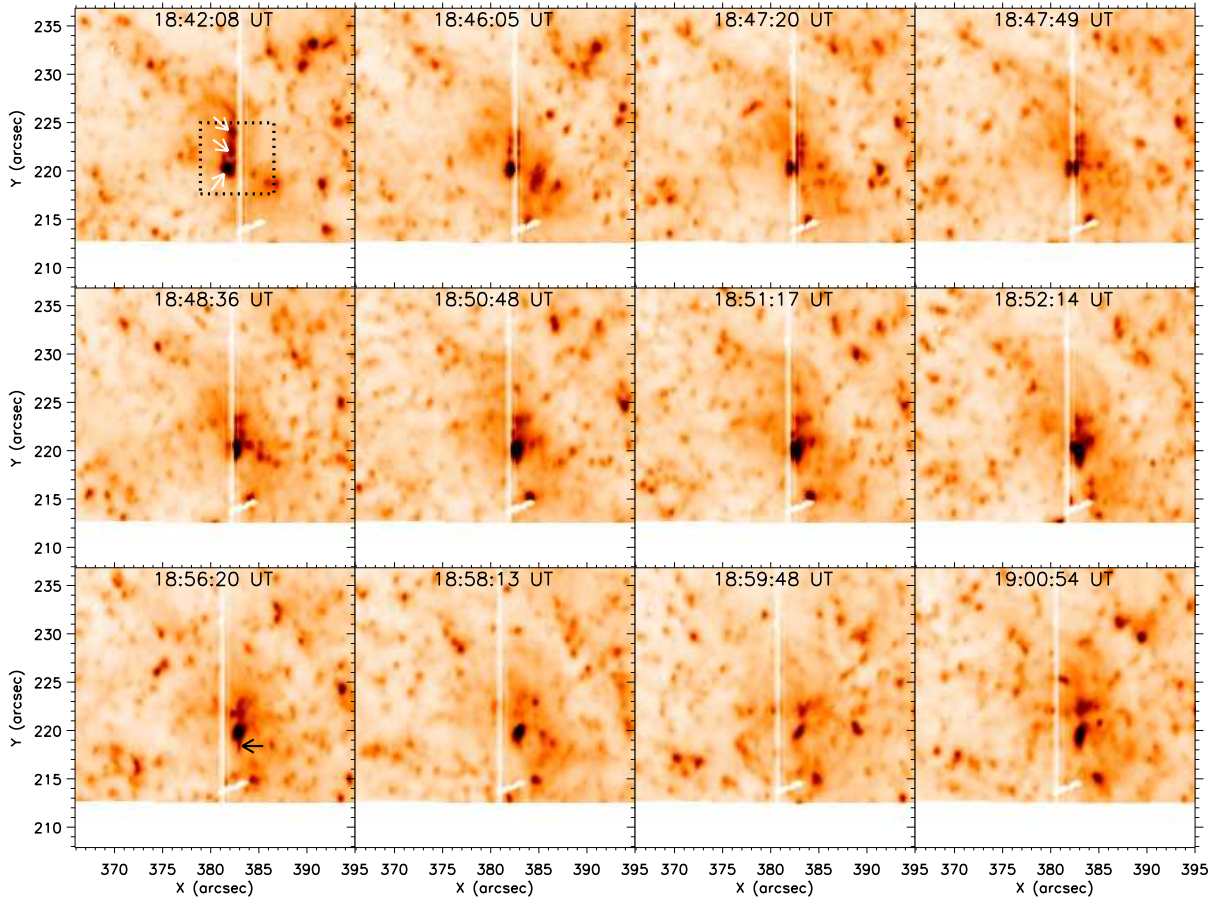}
\caption{The explosive event  in IRIS 1330\,\AA\ slit-jaw images  (in reversed colour). The white vertical line is the location of the IRIS spectral slit. The arrows on the image at 18:42:08\,UT mark the three bright cores discussed in the text, and the dotted-line square outlines the region from which the lightcurves in Fig.~\ref{fig_ee_lcs} are produced. 
The arrow on the image at 18:56:20~UT denotes a jet-like feature.\label{fig_ee_sjw}}
\end{figure*}

IRIS provides a unique opportunity to study solar phenomena in simultaneously taken spectroscopy and imaging data.   Fig.\,\ref{fig_ee_sjw} displays the evolution of the event seen in the 1330\,\AA\ slit-jaw filter. At the beginning of the observations (18:42:08\,UT), a brightening system with three bright cores can be seen in the area (see the  arrows in the first panel of Fig.\,\ref{fig_ee_sjw}). The EE is distinctly present in the AIA UV 1600\,\AA\ and 1700\,\AA\ channels with the same general shape. The three cores, however, are impossible to distinguish due to the lower  (up to 3--4 times) spatial resolution of AIA.   The event is seen in the AIA 1600\,\AA\ images  from 17:54\,UT to 19:19\,UT, i.e. during 85~mins. 

\par
\begin{figure*}
\includegraphics[width=15cm,clip,trim=0.5cm 2cm 0.1cm 0cm]{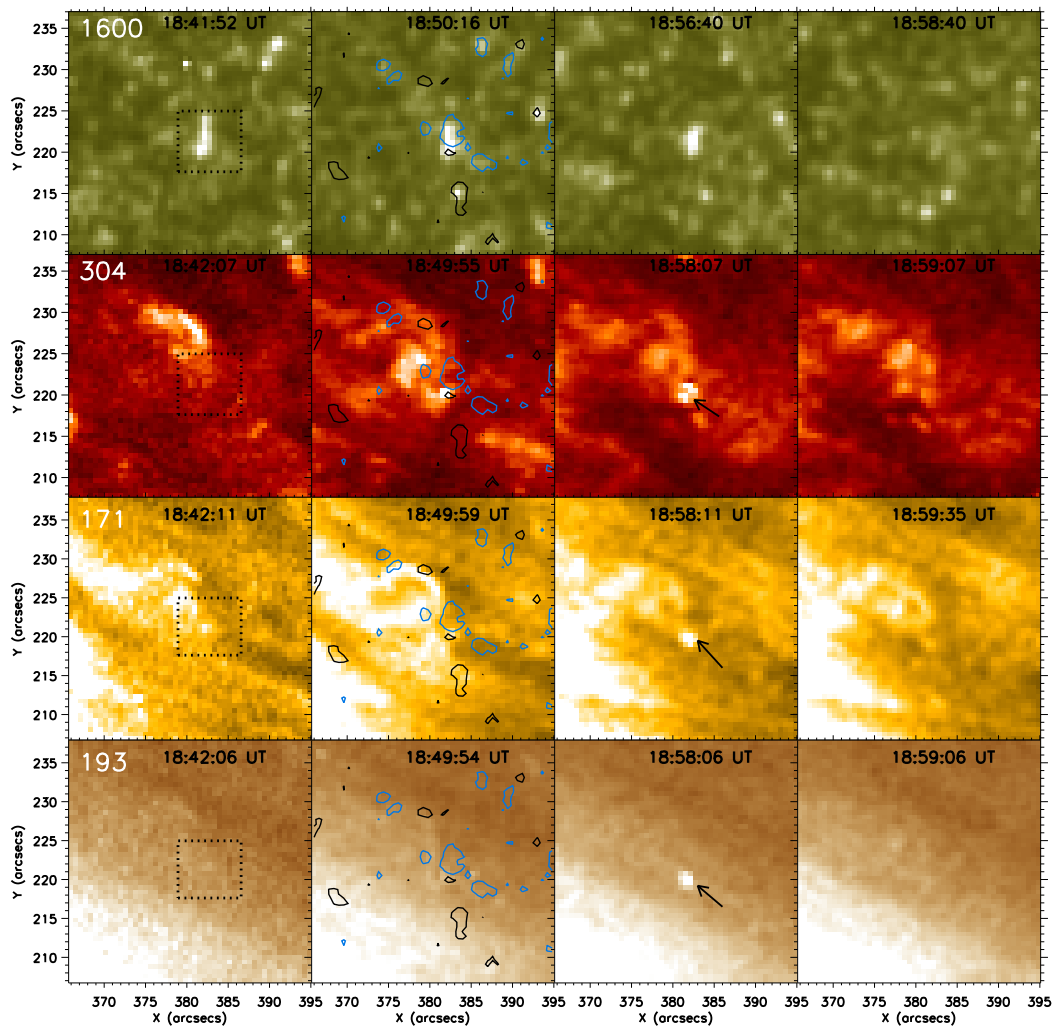}
\caption{The explosive event  in the AIA channels. {\bf From top to bottom}: 1600\,\AA, 304\,\AA,  171\,\AA, and 193\,\AA. The dotted lines on the first column outline the region from which the lightcurves (Fig.\,\ref{fig_ee_lcs}) are calculated. The contour of the magnetic flux density is over-plotted on the second column of the images (black: $-$20\,Mx\,cm$^{-2}$, cyan: 20\,Mx\,cm$^{-2}$). The arrows in the third column of AIA 304\,\AA, 171\,\AA\ and 193\,\AA\ images denote the brightening associated with the EE. 
Full cadence of the images (including 211\,\AA\ and 131\,\AA\ channels) is given as online animation.\label{fig_ee_aia}}
\end{figure*}

The event was also investigated in the 304\,\AA, 171\,\AA\ and 211\,\AA\ channels shown in Fig.\,\ref{fig_ee_aia}.  In the AIA~304~\AA\ and 171~\AA\  images a dynamically evolving bright complex feature is present. During its evolution some loop pattern becomes apparent, but to clearly identify individual loops and to link to the underlying magnetic field configuration is close to impossible. One of the  footpoints of the complex brightening is rooted in the three-core 1600~\AA\ feature where also the EE is found to originate. The investigation of the associated HMI magnetograms shown in Fig.\,\ref{fig_ee_hmi} indicates that the EE is located above two opposite-polarity cancelling magnetic features (see the online animation Fig.~\ref{fig_movie1}). It  also shows that the HMI resolution is too low to make any reliable field extrapolations for  more detailed investigation of   the magnetic field structures and associated small-scale dynamics. The bright core feature linked to the EE  appears very dynamic in the IRIS 1330\,\AA\ images and a very close look reveals that continuously small-scale ejections take place during the entire IRIS observation period. Most of the ejections have a mini-jet-like shape. They propagate in random directions, and  some of them (as the clear example shown with a black arrow in image at 18:56:20\,UT of Fig.\,\ref{fig_ee_sjw}) have a width of one pixel (i.e. 0.166\arcsec). Such fine scale jet structures have so far only being seen  with ground-based telescopes, e.g. Ellerman bombs \citep[e.g.][]{2011ApJ...736...71W}. Part of the ejected material appears to contract back to the source region. Towards the end of the 1330\,\AA\ image series a disk projection of what appears to be a small cloud-like feature is ejected after which the source region become weaker. To follow this evolution please view the online close-look  SJ  1330\,\AA\ image animation (Fig.\,\ref{fig_movie2}). \citet{1998ApJ...504L.123C} found an  association between  EEs observed in Si~{\sc iv}~1402~\AA\ and chromospheric upflows identified as  blue-shifted \halpha\ profile at 1\arcsec\ spatial resolution. No jet features were observed but rather ``dark dots'' with a size of 2\arcsec--3\arcsec\ and lifetime of 1--2 min were identified in the Dopplergram with an upflow velocity of 15--30~\kms. The authors suggested that the ``chromospheric upflow events may be the manifestation of cool plasma material flowing into magnetically diffusive regions, while explosive events represent hot plasma material flowing out of the same regions''.

\par
\begin{figure*}
\includegraphics[width=0.5\textwidth,clip,trim=0cm 1.2cm 0cm 0cm]{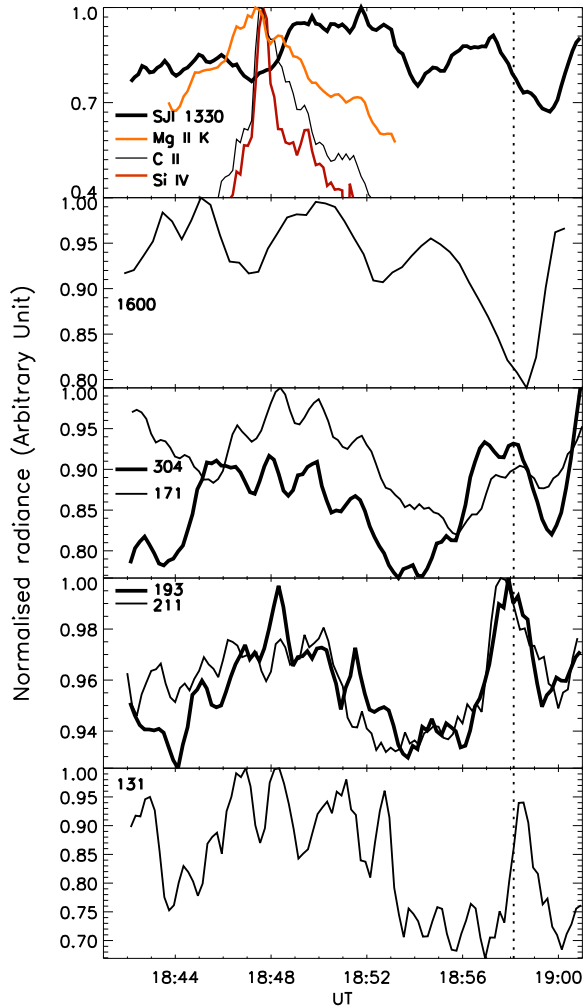}
\caption{IRIS and AIA lightcurves of the region outlined by dotted line in Figs.\,\ref{fig_ee_sjw} and \ref{fig_ee_aia}. The lightcurves of the spectral lines are produced by integrating over the whole line during the time interval when the slit was crossing the event. The vertical dotted line denotes the time from which the emission of AIA 171\,\AA, 193\,\AA, 211\,\AA\ and 131\,\AA\ is used to calculate the EM Loci (Fig.\,\ref{fig_aia_emloci}). \label{fig_ee_lcs}}
\end{figure*}

In Fig.\ref{fig_ee_lcs}, top panel, we present the  lightcurves in the three spectral lines together with the lightcurve of the event  in the SJ images (see the remarks in Fig.\,\ref{fig_ee_sjw}). The lightcurves of  the emission in the AIA EUV channels are shown in the panels below and are obtained  from the boxed region over-plotted on  the first column in Fig.\,\ref{fig_ee_aia}.  The region selected  for producing the lightcurves is larger  (7.65\arcsec$\times$7.35\arcsec)  in comparison to the event because we need to account for the movement of the feature and for the activity in its close proximity  (perhaps triggered by the EE), see Fig.\,\ref{fig_ee_sjw}. Thus the lightcurves are formed  partially by the EE source region  and also by the loops rooted in it. The lightcurves are smoothed by 3 frames.  We need to point out that the dip in the SJ lightcurve during the peak of the event seen in the spectral lines is purely instrumental (see Sect.\,\ref{sect_obs} for more details). The slit is obscuring the event during this period of time and because of the small size of the event at least 1/3 of its emission is blocked. 

\par
 The lightcurves in Fig.\ref{fig_ee_lcs} show two periods of brightening increase.  The first lasts for $\sim$8~min but is actually composed of several intensity peaks each with duration between 90~s  and 120~s.   The second brightening has the same duration as the individual spikes of the first brightening.  These intensity variations are clearly linked to the small-scale ejections and brightness increase in the source region. The first intensity increase starts in the AIA~304~\AA\ channel at $\sim$18:44~UT and ends at $\sim$18:53~UT. The start in the SJ images cannot be defined because of the spectroscopic observations. In the 171~\AA\ channel the flaring of the EE begins later, i.e. $\sim$18:46~UT, and stops earlier at $\sim$18:52~UT. The event is first seen in the AIA 193~\AA\  channel at the same time as in AIA~304~\AA. In AIA~131~\AA, it appears to be delayed by one minute with respect to the 304~\AA\ and 193~\AA\ channels but this can also be due to the weak signal in this channel. The start of the brightening in the 211~\AA\ channel is impossible to determine. 

\par
 At 18:55\,UT, the EE starts to flare up again (only seen in the imaging data) and reaches maximum in the IRIS 1330\,\AA\ slit-jaw images at 18:56:01\,UT. During this stage, the event shows a clear jet-like structure seen in IRIS 1330\,\AA\ SJ images (black arrow in image at 18:56:20\,UT of Fig.\,\ref{fig_ee_sjw}), and can also be followed in the AIA UV/EUV channels (arrows in the 3rd column of Fig.\,\ref{fig_ee_aia}). The time of the emission maxima are 18:56:55\,UT in 304\,\AA, 18:57:47\,UT in 171\,\AA\ (44~s later with respect to AIA~304~\AA), 18:57:42\,UT in 193\,\AA\ (102~s), 18:57:47\,UT in 211\,\AA\ (102~s), and  18:58:32\,UT for 131\,\AA\ (152~s). Because of the 12~s cadence of the AIA data, the 171~\AA, 193~\AA, and 211~\AA\ actually respond simultaneously. Only the 131~\AA\ channel is delayed with respect to the other coronal temperature channels by $\approx$50~s. The emission recorded in the 131~\AA\ channel is known to be dominated by Fe~{\sc viii} and several transition region lines.  Therefore, the later response can be related to a cooling rather than heating during the EE. The 171~\AA\ channel is dominated by Fe~{\sc ix} but if a feature at low temperature is observed, the recorded emission will be at temperatures logT (K) $<$ 5.7  \citep{2012SoPh..280..425V, 2011ApJ...730...85B}. Although the AIA~193~\AA\ channel is dominated by three Fe~{\sc xii} lines, it  has a significant transition-region emission contribution from non-identified lines \citep{2011A&A...535A..46D}.  The AIA 211~\AA\ channel is also known to record transition region emission, a lot coming from Fe~{\sc viii} lines.  However, 50~\%\ of the lines emitting in this channel remain unidentified \citep{2011A&A...535A..46D}. Only in active regions we do have a strong contribution from Fe~{\sc xiv}.  To conclude, the AIA channels do not give a straight answer on the temperature of the observed explosive event. 
 
 \par
\begin{figure*}
\includegraphics[clip,trim=1cm 0.3cm 0.5cm 0.5cm]{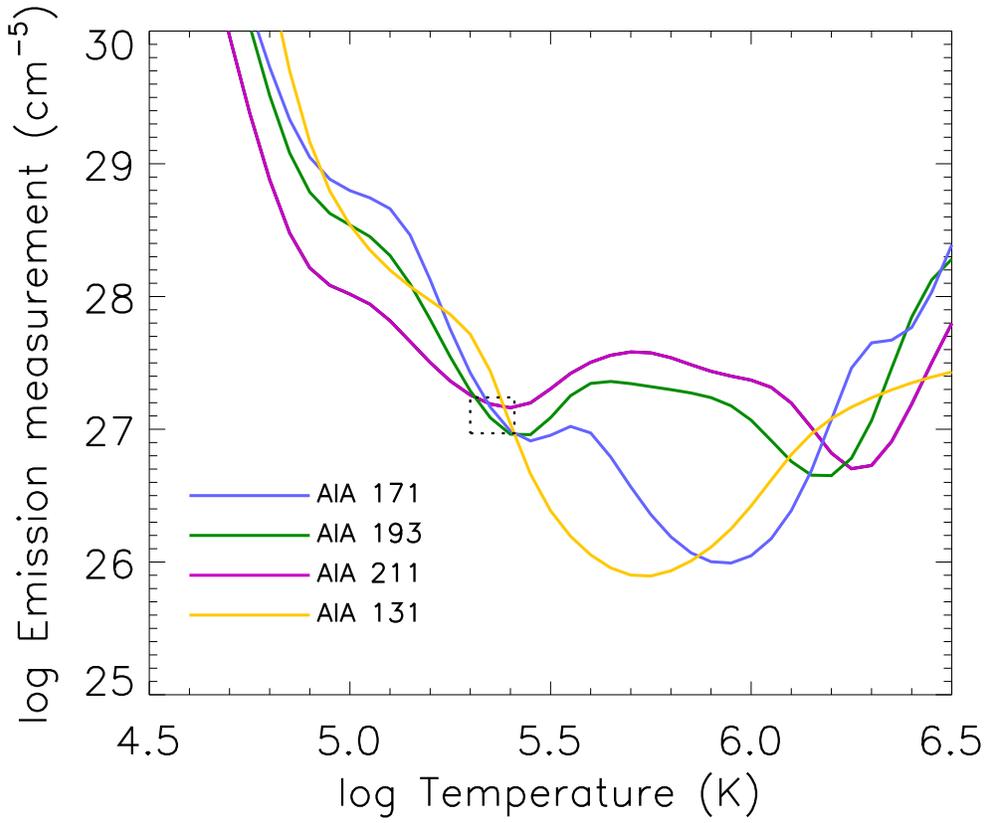}
\caption{EM loci curves made using the AIA 171\,\AA, 193\,\AA, 211\,\AA\ and 131\,\AA\ channels  taken around 18:58\,UT. The dashed-line square  marks the crossing  of the curves at  log T = $5.36\pm0.06$\,K. \label{fig_aia_emloci}}
\end{figure*}

The clear intensity increase in  AIA 171\,\AA, 193\,\AA, 211\,\AA\ and 131\,\AA\ suggests that during the explosive event, plasma was ejected reaching temperatures to which all of these channels are sensitive to.  To determine this temperature, we used  the emission measure (EM) Loci method \citep[see, e.g. ][etc.]{2011ApJ...740....2W, 2013ApJ...775L..32A}. The EM loci curves are constructed from the channel response functions calculated using the method described in \citet{2011A&A...535A..46D} and the emission at around 18:58\,UT. The resulting curves  are shown in Fig.\,\ref{fig_aia_emloci}. The EM loci suggests that log\,T\,= $5.36\pm0.06$\,(K) is the most probable temperature of the event. The second loci-curve clustering at 6.18$\pm$0.1~K is less plausible following the discussion above on the temperature response of the AIA EUV channels. An EUV active-region jet was reported by \citet{2003ApJ...584.1084C} in TRACE~1600~\AA\ (transition-region and continuum emission) and 171~\AA. The jet, which was much larger than the jets reported here, had a temperature of 2--3$\times$10$^5$~K. The temperature of the jet was obtained from a TRACE filter ratio method.

\subsection{Magnetic field of the explosive event}
\label{sect_mag}

\begin{figure*}
\includegraphics[width=17cm,clip,trim=0.1cm 2cm 0.1cm 0cm]{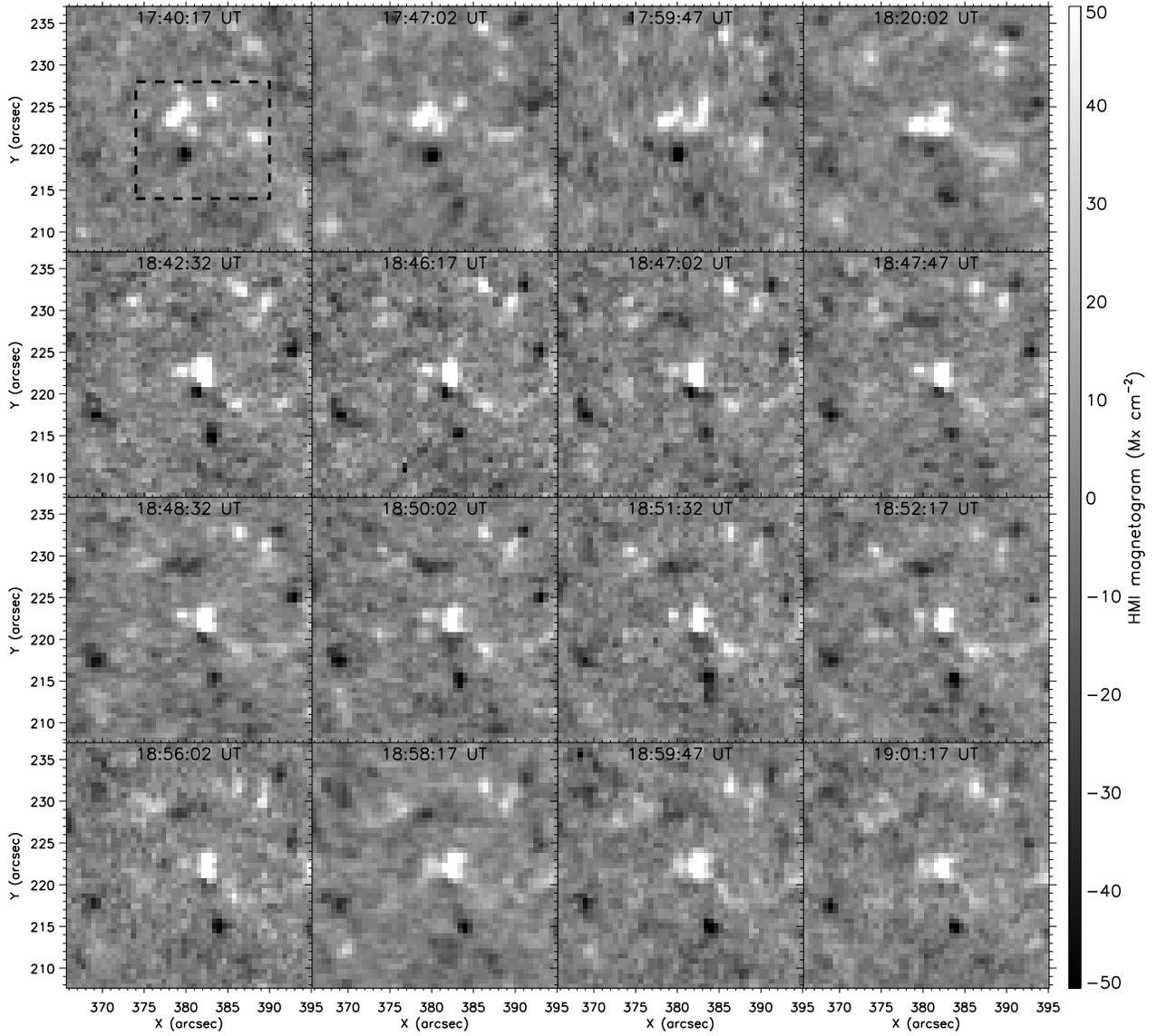}
\caption{HMI magnetograms of the explosive event. The over-plotted dashed line square is the region from where the magnetic flux is calculated in Fig.~\ref{fig_ee_blcs}.
\label{fig_ee_hmi}}
\end{figure*}

Having now the great opportunity provided by IRIS to observe spectroscopically the Sun at sub-arcsecond resolution, we face the challenge of  having both incompatible  coronal imaging and also magnetic field data.  Nevertheless, we analysed in detail the only available data from HMI.
 After the magnetic field associated with the explosive event was identified, we tracked its evolution starting an hour before the IRIS observations. A selection of HMI images is shown in Fig.\,\ref{fig_ee_hmi} and the image animation in the online material (Fig.~\ref{fig_movie1}).  At the start of the selected dataset (i.e. 17:40~UT), a bipolar region is found in the area of the EE with one negative polarity fragment and a few positive ones. The distance between the two closest negative and positive fragments is $\sim$2\arcsec. The positive  and negative fragments are moving towards each other while at the same time the positive flux increases which is either due to  flux emergence or convergence (the  spatial scale and sensitivity of HMI does not permit a judgement as to which one of the two mechanisms is at work). Magnetic flux cancellation  associated with  EEs has been previously reported  by \citet{1998ApJ...497L.109C} and \citet{2008ApJ...687.1398M}. In both studies, however, only around 63\% and 19\%  (103 out of 165 and 7 out of 37),   of the EEs resulted from flux cancelation, respectively. As in the present case, instrumental limitations could be the reason for this result and possibly larger or even all EEs are associated with magnetic flux cancellation. Thus studies using higher magnetic field resolution and sensitivity data are required, as it will be discussed further in the text. Jet-like phenomena of various sizes and temperatures have been related to magnetic flux cancellation  \citep[e.g.,][]{1999ApJ...513L..75C, 2003ApJ...584.1084C, 2008A&A...491..279C, 2012A&A...545A..67M, 2012A&A...548A..62H, 2014ApJ...783...11A}. The authors of all these studies  suggest magnetic reconnection as the most probable driving mechanism.

\par
\begin{figure*}
\includegraphics[width=8cm,clip,trim=1cm 0.5cm 0.5cm 0.5cm]{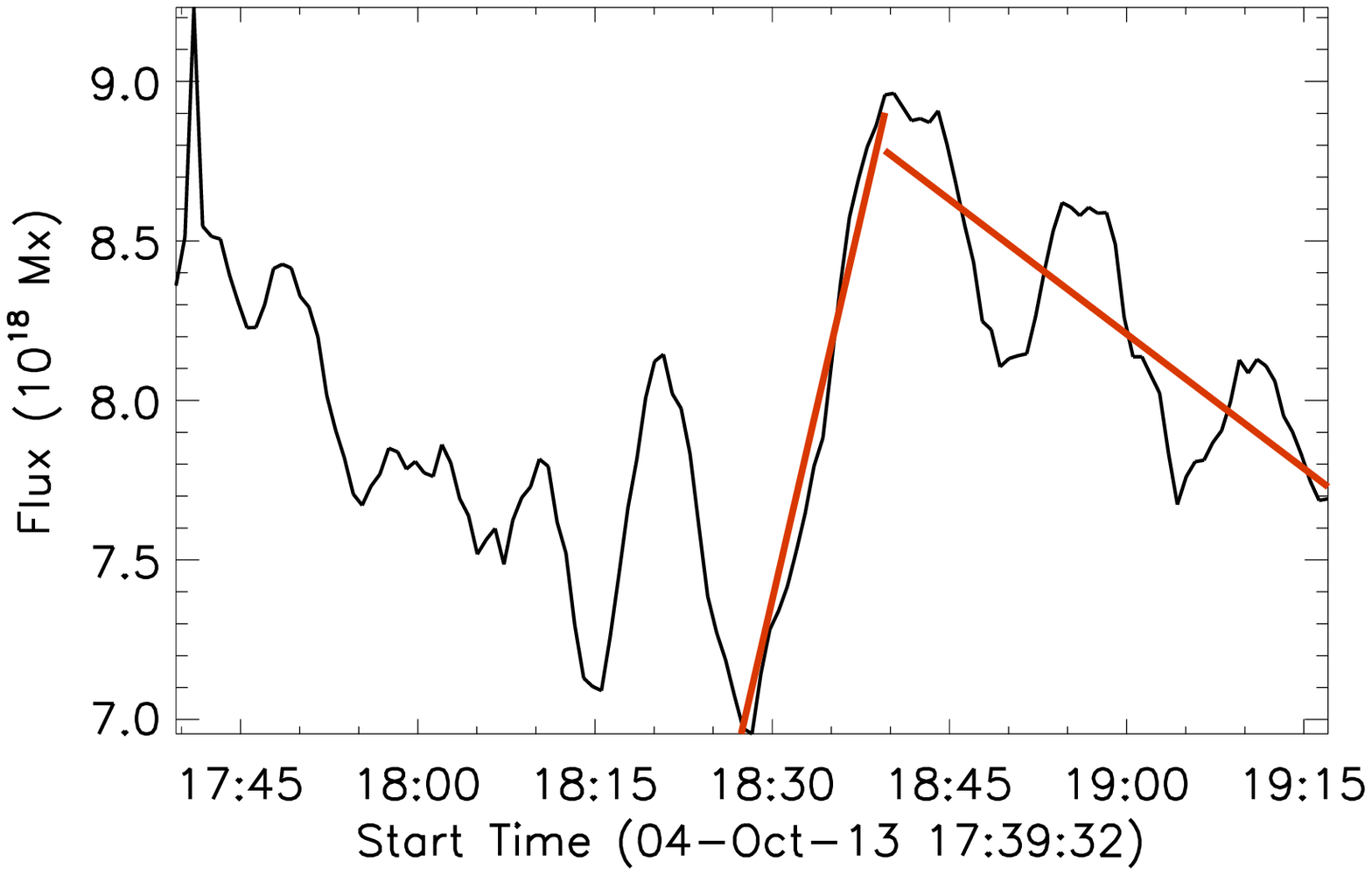}
\includegraphics[width=8cm,clip,trim=1cm 0.5cm 0.5cm 0.5cm]{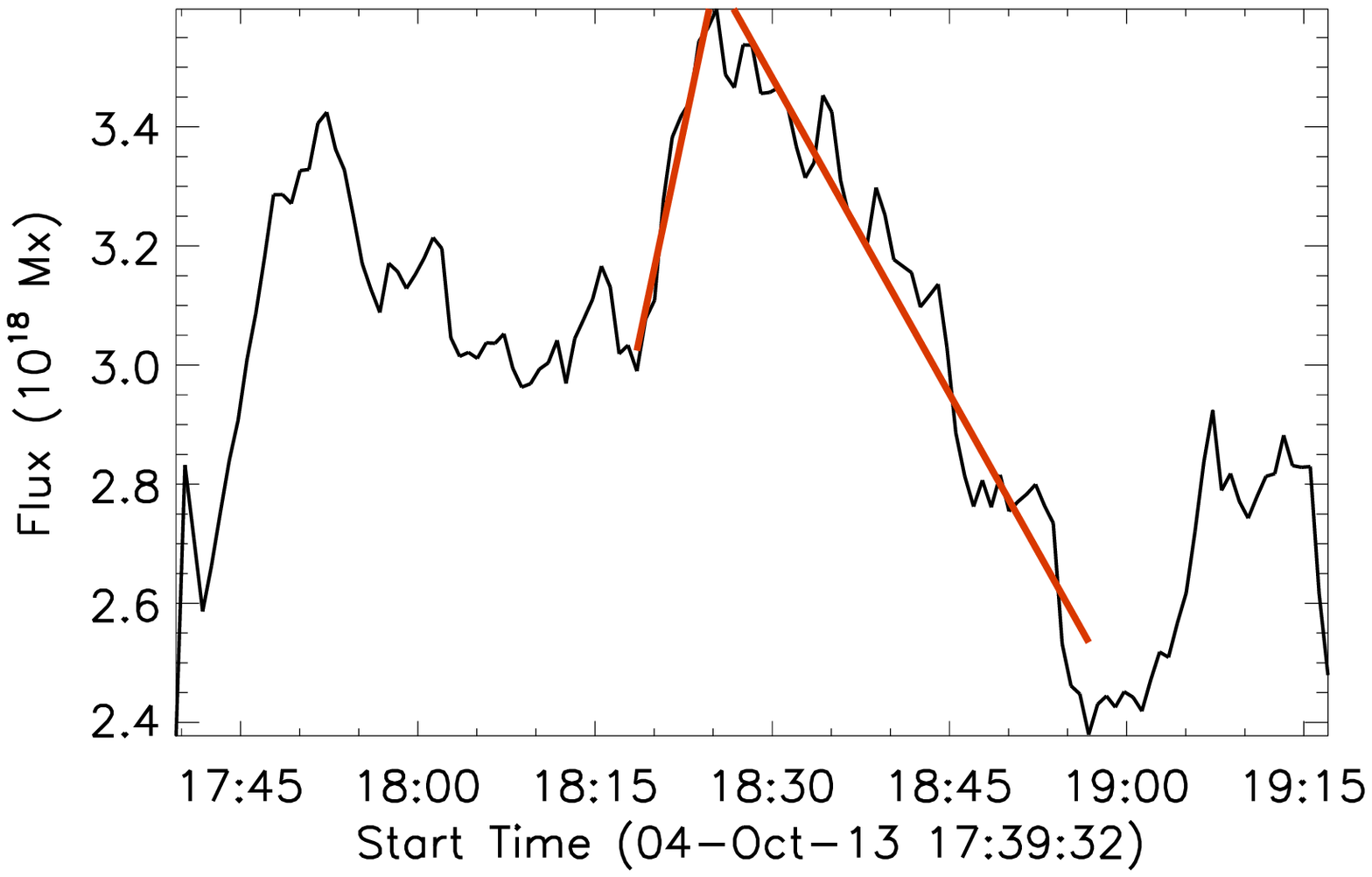}
\caption{Temporal variations of the positive (left) and negative (right) magnetic fluxes. The red lines represent  the linear fit to the period of flux emergence and cancellation.\label{fig_ee_blcs}}
\end{figure*}
\begin{figure}
\epsscale{.80}
\plotone{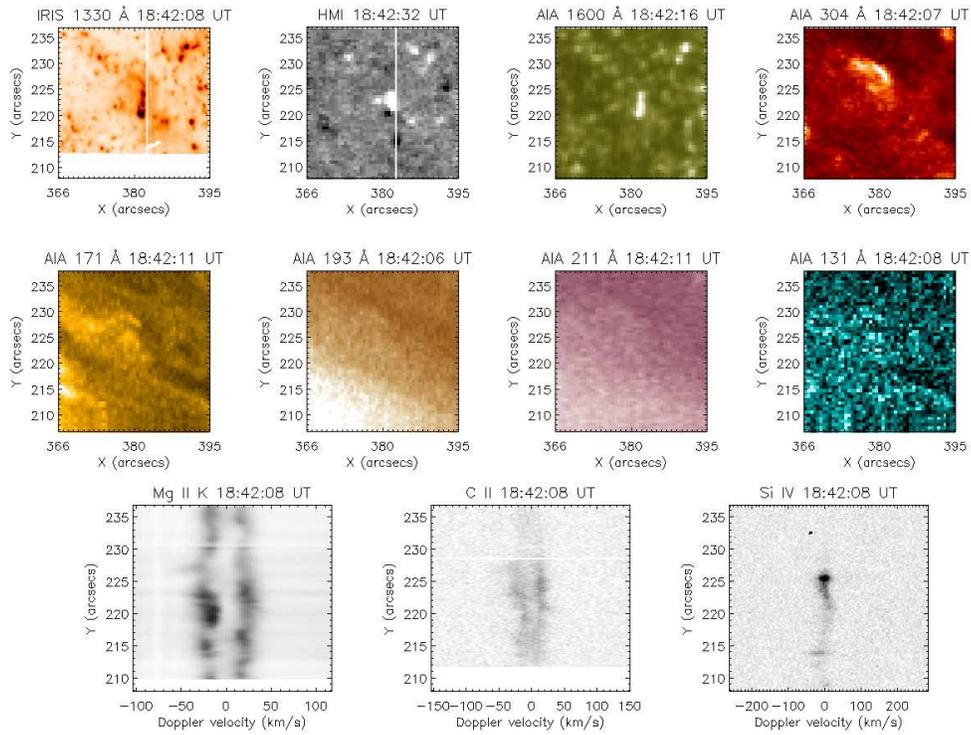}
\caption{A frozen frame of the animation during the explosive event. {\bf Top row}: IRIS 1330\,\AA\ SJ images (in reversed colour table), HMI with over-plotted the IRIS slit position (white vertical line), AIA 1600\,\AA, and AIA 304\,\AA\ images; middle row: AIA 171\,\AA, AIA 193\,\AA, AIA 211\,\AA, and AIA 131\,\AA\ images. {\bf Bottom row}: slit images of  the three spectral lines, \mgiik, \cii\ and \siiv.%Higher resolution animation can be requested from the addressed author.
\label{fig_movie1}}
\end{figure}
\begin{figure}
\epsscale{.80}
\plotone{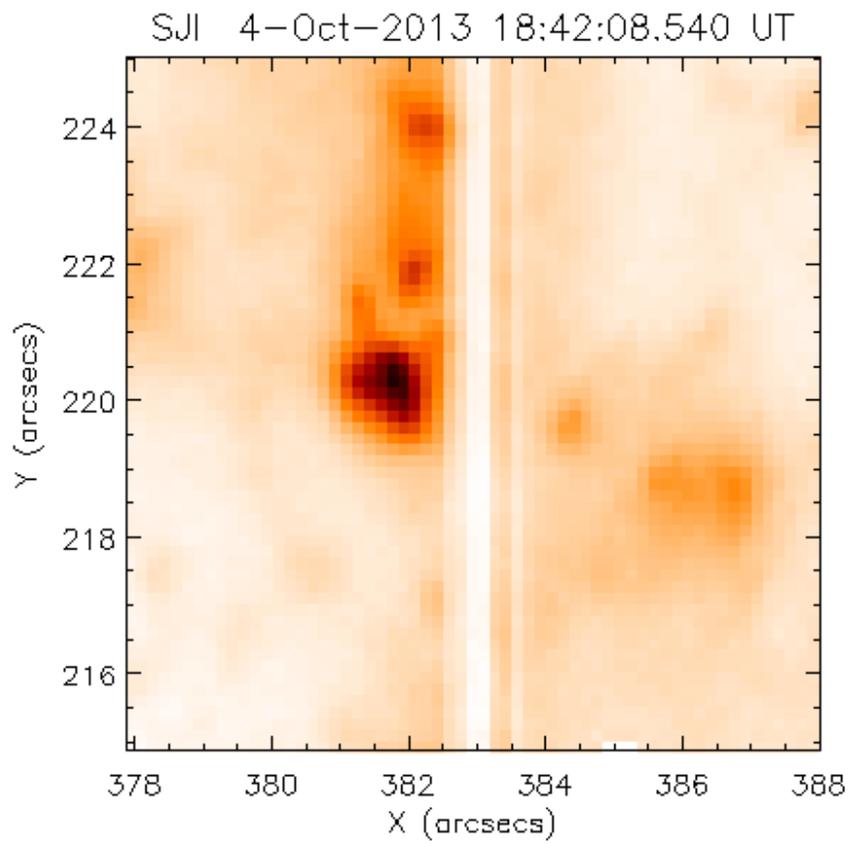}
\caption{A frozen frame of the animation of the explosive event closely viewed in IRIS 1330\,\AA\ SJ images.\label{fig_movie2}}
\end{figure}

To evaluate the magnetic activity during the EE, we produced the temporal variations of the total positive and negative magnetic flux  from the boxed region in Fig.\,\ref{fig_ee_hmi}. A limited field -of-view was used around the region of the EE.  The temporal variations shown in Fig.\,\ref{fig_ee_blcs} reveal that a flux increase starts prior to the event in both the positive (from $\sim$18:27~UT until 18:43~UT) and the negative (from 18:18~UT until 18:25~UT) polarities.  The increase of the positive flux is at a rate of $2.7\times10^{15}$\,Mx\,s$^{-1}$, while the increase of the negative is  $1.6\times10^{15}$\,Mx\,s$^{-1}$. At  the beginning of the IRIS observations at 18:42\,UT, the distance between the negative and  positive magnetic fragments is within one HMI pixel when the magnetic-field cancellation was already ongoing. The cancellation rate is $4.7\times10^{14}$\,Mx\,s$^{-1}$ for the positive flux (linear fit to the curve between 18:39\,UT and 19:17\,UT), and $5.9\times10^{14}$\,Mx\,s$^{-1}$ for the negative flux  (linear fit to the curve between 18:25\,UT and 18:59\,UT). We should stress here that these estimations are only a low limit because of instrumental limitations. The cancellation rates are close to those found for an X-ray jet \citep{2012A&A...548A..62H}, and much smaller than those recorded during a GOES C4.3 flare \citep{huangzh2013}. As discussed in \citet{2012A&A...548A..62H}, this flux cancellation rate strongly suggests that the event observed here results from an impulsive energy release most possibly magnetic reconnection. To compare, \citet{2002SoPh..207...73C} found $3.6\times10^{14}$\,Mx\,s$^{-1}$ and $9.7\times10^{14}$\,Mx\,s$^{-1}$ of flux cancellation rate for two cancellation sites, which the authors suggests to be consistent  with a Sweet-Parker magnetic reconnection model. In the present case, the combined changes of the locations and the flux of different patches will result in reconfiguration of the field line connectivity that can only take place through magnetic reconnection. To fully understand the dynamical evolution of this event, reliable 3D models of the magnetic field are required. With the small size of this event, which represents only a few HMI pixels, a representation of the small-scale magnetic field is not possible to obtain by any extrapolation models. Without this, a detailed understanding of the underlying mechanism that drives this event is very challenging.

\section{Conclusions}
\label{sect_concl}
The real nature of explosive events is still under debate. Jets produced by magnetic reconnection, siphon flows in small-scaled loops and swirling jets have been proposed as the phenomena causing explosive-event line profiles.  Please note that none of these interpretations was confirmed by imaging information until present. Here, we analyse an explosive event witnessed by strong non-Gaussian profiles with up to 150\,\kms\ blue- and red-shifted components in the \siiv line. The EE was associated with a small ($\sim$1.5\arcsec) compact bright-point-like structure in the IRIS unprecedented high-spatial, temporal and spectral resolution observations combining them with imaging and magnetic field data from AIA and HMI, respectively. We found, for the first time,  that an ``explosive event'' phenomenon is associated with continuous small-scale plasma ejections  and retractions on a sub-arcsecond scale (jet's width of 0.166\arcsec\ or  120 km) observed in the solar chromosphere (IRIS SJ 1330\,\AA).

\par
In the AIA 304\,\AA\ and 171\,\AA\ channels, the explosive event appears to be located in the footpoints of a  complex multiple loop system, which is also confirmed by the HMI magnetograms. Magnetic flux emergence or convergence followed by flux cancellation at the location of the explosive event was found, suggesting that magnetic reconnection producing high velocity plasma outflows was taking place. Brightenings observed in the AIA 304~\AA, 171~\AA\, 211~\AA\ and 131~\AA\ channels with duration between 90~s and 120~s are  most probably produced by the plasma ejections (also responsible for the explosive event) seen in the SJ 1330~\AA\ images.

\par
The \mgiik\ and \cii\  lines observed by IRIS show self-reversed profiles, with the explosive event affects the line wings (i.e., the emission peaks) but not the absorption dips (the line centres). The wings of the  \cii\ line,  above 50~\kms\ behave as  the \siiv\ line wings, suggesting the wings of these two lines are formed at similar temperatures while the wings of \mgiik\ are affected by both Doppler-shift and emission contribution possibly related to their formation in the lower chromosphere. A predominantly strong red-shifted emission is observed in all lines  at one edge (west) of the small-scale (~1.5\arcsec) bright structure (as  seen in the SJ images) while at the opposite edge (east) the emission is predominantly blue-shifted suggesting a down and up-flow, respectively, on a scale as low as  0.16\arcsec (or 118 km). A particular feature was observed  at the west edge of the SJ bright feature, where the \mgiik\ and \cii\ lines show profiles with a strong emission increase in the red peak but blue-shifted towards the line centre with the opposite being observed in the east edge (i.e. blue wing increase shifted towards the line centre). Because of the lack of radiative transfer calculation in the presence of very energetic events, to speculate of what causes these profiles is not presently possible. This bring challenge to radiation transfer calculation of these two optically thick lines.

\par
RB asymmetry analysis of the  \mgiik, \cii\  and \siiv\, suggests that the flows (jets) producing the explosive event originate in the low chromosphere. The plasma up-flows clearly dominate the down-flows which indicates that plasma heated to high temperatures is ejected and after depositing energy in the high transition region or corona falls back to the chromosphere. The temperature of the event derived using the EM Loci method is log\,T\,$=5.36\pm0.06$\,(K). The phenomena associated with the EE, therefore, directly contribute to the heating of the solar transition region. Their impact on coronal heating is still to be investigated. As we mentioned in the introduction, the only study that has concluded on the direct coronal contribution of EEs \citep{2002A&A...392..309T}  is based  only on two events.

\par
Although our case study can not answer all the questions listed in the introduction, it puts our knowledge a step forward by demonstrating one of the possible phenomena producing EEs.  The present study  provides a wealth of information on the behavior of chromospheric lines in non-equilibrium ionization. A future forward chromospheric \cii\ and \mgiik, and transition region \siiv\ modelling studies \citep[including non-equilibrium ionisation effects, see][]{2013A&A...557L...9D} can then provide important clues on the physical mechanism(s) in action during phenomena witnessed by explosive events. 

\par
The  temporal behaviour of the chromospheric and transition-region lines clearly demonstrates that the scenario suggested by  \citet{1999ApJ...513L..75C} where chromospheric upflows  represent cool plasma material flowing into magnetically diffusive regions, while explosive events represent hot plasma material outflow from the reconnection site, is not valid in the present case. The imaging data on the other hand show that the plasma up- and down-flows are produced by plasma ejection and retraction rather than bi-directional reconnection outflows  as suggested by \citet{1991JGR....96.9399D} and  \citet{1997Natur.386..811I}. The present phenomenon does not support the scenario of a swirling upflow \citep{2011A&A...532L...9C} as a possible phenomenon witnessed by EEs. We would like to stress that although our case study does not endorse these interpretations, they still remain  plausible explanations for other EE-associated phenomena. To really extend our knowledge on small-scale events using high-resolution IRIS observations, we need in parallel much higher resolution magnetograms than those are currently available.

\par
A just started statistical study covering events in various regions on the Sun should provide a lot of the missing pieces of the puzzle ``explosive event''. New recently obtained magnetic field and \halpha\  observations from the Swedish Solar Telescope together with IRIS may also shed more light on how exactly the small-scale jets producing explosive events are generated.

\acknowledgments
{\it Acknowledgments:}
This research is supported by the China 973 program 2012CB825601, and the National Natural Science Foundation of China under contract 41274178, 41404135. Research at the Armagh Observatory is grant-aided by the N. Ireland Department of Culture, Arts and Leisure. We thank STFC (grant ST/J001082/1) and the Leverhulme Trust for the financial support. We thank the anonymous referee for the critical and constructive comments. We thank Dr. Hui Tian for many useful discussions. IRIS is a NASA small explorer mission developed and operated by LMSAL with mission operations executed at NASA Ames Research centre and major contributions to downlink communications funded by the Norwegian Space Center (NSC, Norway) through an ESA PRODEX contract. AIA and HMI data is courtesy of SDO (NASA).

{\it Facilities:} \facility{IRIS}, \facility{SDO/AIA}, \facility{SDO/HMI}.

%1

%2

%3 

%4

%5

%6

%7

%8

%9

%10

%11

%12

%13

%14

\bibliographystyle{aa}
\bibliography{references1}

\appendix

\section{Online material}

\end{document}